\documentclass[showkeys,superscriptaddress,floatfix,prd,10pt,aps,twocolumn]{revtex4-2}
\usepackage{graphicx,epstopdf}
\usepackage[caption=false]{subfig}
\usepackage{appendix}
\usepackage[T1]{fontenc}
\usepackage{lmodern}
\usepackage[dvipsnames,x11names]{xcolor}
\usepackage[colorlinks=true,linkcolor=Magenta!80!black,citecolor=Magenta!70!black,urlcolor=Magenta!80!black]{hyperref}
\usepackage[sort&compress]{natbib}
\usepackage{lipsum}
\usepackage{morefloats}
\usepackage[pdf]{pstricks}
\usepackage{amsmath,amsthm}
\usepackage{amssymb}
\usepackage{amsfonts}
\usepackage{rotating}
\usepackage{cancel}
\usepackage{mathtools}
\usepackage{bbm}
\usepackage{dsfont}
\usepackage{bbold}
\usepackage{multirow}
\usepackage{ulem}
\usepackage{physics}
\usepackage{orcidlink}
\usepackage{colortbl}

\def\pslash{p\!\!\!\slash }
\def\qslash{q\!\!\!\slash }
\def\xslash{x\!\!\!\slash }
\def\eslash{\varepsilon\!\!\!\slash }

\begin{document}

\title{Electromagnetic form factors: A window into the $D\Lambda_c$, $D^*\Lambda_c$, and $D\Lambda_c^*$ molecular structure}

\author{Ula\c{s} \"{O}zdem\orcidlink{0000-0002-1907-2894}}%
\email[]{ulasozdem@aydin.edu.tr}
\affiliation{Health Services Vocational School of Higher Education, Istanbul Aydin University, Sefakoy-Kucukcekmece, 34295 Istanbul, T\"{u}rkiye}


\begin{abstract}
The internal structure of exotic hadrons remains one of the most compelling puzzles in strong interaction physics. In this work, we provide crucial insights into the nature of doubly-charmed pentaquarks by investigating their electromagnetic properties. Using QCD light-cone sum rules, we present the first comprehensive calculation of the magnetic dipole moments of $D\Lambda_c$, $D^*\Lambda_c$, and $D\Lambda_c^*$ molecular pentaquarks with $J^P = \frac{1}{2}^-$, $\frac{3}{2}^-$, and $\frac{3}{2}^-$, respectively. Our analysis reveals a striking hierarchy of magnetic moments: $\mu_{D\Lambda_c^*} > \mu_{D^*\Lambda_c} > \mu_{D\Lambda_c}$, driven by distinct quark-level mechanisms. While light quarks dominate the overall response, we find that charm quark contributions become strategically important when light quark contributions partially cancel. Beyond dipole moments, we predict higher multipoles--electric quadrupole and magnetic octupole moments--for the spin-$3/2$ $D^*\Lambda_c$ and $D\Lambda_c^*$ states, which fingerprint the spatial deformation of these configurations, revealing prolate versus oblate charge distributions.  These results provide the first systematic predictions for electromagnetic moments of molecular pentaquark configurations, establishing essential benchmarks for future theoretical and experimental studies. By systematically comparing our predictions with both compact doubly-charmed and hidden-charm pentaquark configurations, we establish robust benchmarks that discriminate between competing structural models, ultimately resolving the nature of doubly-charmed exotic hadrons.
\end{abstract}

\maketitle

\section{Introduction}\label{motivation}

The traditional quark model does not adequately describe exotic hadrons, particularly compact multiquark configurations and molecular states. During the past three decades, these systems have drawn significant attention in hadron physics and motivated a series of experimental and theoretical investigations. Hidden-charm pentaquarks, such as $\mathrm{P_{c}(4380)}$, $\mathrm{P_{c}(4440)}$, $\mathrm{P_{c}(4457)}$, $\mathrm{P_{c}(4312)}$, $\mathrm{P_{cs}(4459)}$, and $\mathrm{P_{cs}(4338)}$, are among the most prominent examples~\cite{Aaij:2015tga, Aaij:2019vzc, LHCb:2021chn, Collaboration:2022boa}. Since these resonances consist of at least five quarks ($c\bar c uud$ or $c\bar c uds$), they are widely recognized as candidates for hidden-charm pentaquarks. In addition, in 2021 the LHCb Collaboration announced the observation of the first doubly-charmed tetraquark, $T_{cc}^+(3875)$, in the $D^0 D^0 \pi^+$ spectrum~\cite{LHCb:2021vvq, LHCb:2021auc}. Its minimal quark composition is interpreted as $cc\bar u \bar d$, with spin-parity quantum numbers $\rm{J^P} = 1^+$. These experimental findings have triggered a range of theoretical explanations, including compact multiquark scenarios, hadronic molecules, and kinematical effects~\cite{Esposito:2014rxa, Esposito:2016noz, Olsen:2017bmm, Lebed:2016hpi, Nielsen:2009uh, Brambilla:2019esw, Agaev:2020zad, Chen:2016qju, Ali:2017jda, Guo:2017jvc, Liu:2019zoy, Yang:2020atz, Dong:2021juy, Meng:2022ozq, Chen:2022asf, Dong:2021bvy}. Nevertheless, despite the progress achieved, the internal structures and quantum number assignments of exotic hadrons remain one of the most important unresolved issues in modern hadron spectroscopy.

The observation of the $T_{cc}^+(3875)$ tetraquark opens a promising pathway for exploring states beyond conventional hadrons. If this resonance is confirmed as a genuine doubly-charmed tetraquark, it naturally suggests the possible existence of doubly-charmed pentaquarks. This line of reasoning mirrors the arguments that motivated the search for hidden-charm pentaquarks following the discovery of hidden-charm tetraquarks.  Despite the absence of experimental evidence so far, it is important to understand why such pentaquarks have not yet been observed. A variety of theoretical studies have investigated the possible existence, structure, and properties of doubly-charmed pentaquarks ~\cite{Esposito:2014rxa, Esposito:2016noz, Olsen:2017bmm, Lebed:2016hpi, Nielsen:2009uh, Brambilla:2019esw, Agaev:2020zad, Chen:2016qju, Ali:2017jda, Guo:2017jvc, Liu:2019zoy, Yang:2020atz, Dong:2021juy, Meng:2022ozq, Chen:2022asf, Dong:2021bvy, Yang:2025aer}. These works generally indicate that such states are likely located below the corresponding meson–baryon thresholds, which may allow them to form bound states. In comparison with doubly-charmed baryons, pentaquarks carrying two charm quarks are expected to be heavier. However, the complex internal dynamics of multiquark configurations could reduce their masses, thereby making it difficult to distinguish them from ordinary baryons based solely on mass considerations. For this reason, investigating additional properties—such as their electromagnetic structure—could offer valuable insight into the internal composition of doubly-charmed pentaquarks.

While the mass spectra and decay properties of such putative doubly-charmed pentaquarks have been explored, their electromagnetic properties remain largely unexplored. These properties are highly sensitive to the internal quark-gluon dynamics and could provide crucial discrimination between molecular and compact structural models. This work aims to fill this gap by providing the first QCD light-cone sum rule calculation of the magnetic dipole moments for the $D^{(*)}\Lambda_{c}^{(*)}$ molecular pentaquarks.  
The magnetic dipole moment is a fundamental electromagnetic observable that plays an essential role in photo- and electro-production processes, significantly influencing both total and differential cross-sections. As such, it serves as a clean probe of hadronic structure. Compared with mass spectra, production mechanisms, quantum numbers, and decay properties, the electromagnetic characteristics of hadronic states have been less extensively investigated, either experimentally or theoretically. Nevertheless, in recent years, they have attracted increasing attention as key quantities for clarifying the internal structure of exotic states whose nature remains under debate~\cite{Wang:2016dzu,  Ozdem:2025jda, Ozdem:2024rch, Ozdem:2024rqx, Ozdem:2023htj, Ozdem:2022kei, Ozdem:2018qeh, Ortiz-Pacheco:2018ccl, Xu:2020flp, Ozdem:2021btf, Ozdem:2021ugy, Li:2021ryu, Wang:2023iox, Gao:2021hmv, Guo:2023fih, Ozdem:2022iqk, Wang:2022nqs, Ozdem:2022vip,Wang:2022tib, Ozdem:2024jty, Li:2024wxr, Li:2024jlq, Ozdem:2024yel,  Mutuk:2024ltc, Mutuk:2024jxf, Mutuk:2024ach, Ozdem:2025ncd, Ozdem:2025fks, Ozdem:2024suc}. Motivated by these considerations, in this work, we employ the QCD light-cone sum rules to investigate the magnetic dipole moments of the $D^{(\ast)} \Lambda_c^{(\ast)}$ states, assuming that these pentaquarks correspond to the $D \Lambda_c$, $D^{\ast} \Lambda_c$, and $D \Lambda_c^{\ast}$ molecular configurations with quantum numbers $\mathrm{J^P= \tfrac{1}{2}^-}$, $\mathrm{J^P= \tfrac{3}{2}^-}$, and $\mathrm{J^P= \tfrac{3}{2}^-}$, respectively.  
Among the various methods available for evaluating nonperturbative parameters, the QCD light-cone sum rule approach is known for its effectiveness~\cite{Chernyak:1990ag, Braun:1988qv, Balitsky:1989ry}. Within this framework, the relevant correlation function is analyzed on two complementary sides: the \textit{phenomenological side} and the \textit{theoretical side}. By invoking the principle of quark--hadron duality, the results from these two representations are matched, leading to the derivation of sum rules for the magnetic dipole moments at the hadronic level.

The structure of this paper is as follows. In Sec.~\ref{formalism}, we derive the theoretical framework for the magnetic dipole moments of the $D^{(\ast)} \Lambda_c^{(\ast)}$ pentaquarks. The corresponding numerical results are presented in Sec.~\ref{numerical}. Finally, the manuscript concludes with a summary in Sec.~\ref{summary}.

\begin{widetext}
 
\section{Electromagnetic structure and QCD light-cone sum rules}\label{formalism}

The magnetic dipole moments of the spin-1/2 and spin-3/2 $D^{(\ast)} \Lambda_c^{(\ast)}$ pentaquarks ($\mathrm{P_{cc}} \rightarrow J^P = \frac{1}{2}^-$, $\mathrm{P_{cc}^\ast} \rightarrow J^P = \frac{3}{2}^-$) can be evaluated using QCD light-cone sum rules. By analyzing the corresponding correlation functions in a weak external electromagnetic field, denoted by $\gamma$, one can isolate contributions from both perturbative and nonperturbative QCD effects and thus determine the magnetic dipole moments of these states.
 \begin{align} 
 \label{edmn01}
\Pi(p,q)&=i\int d^4x e^{ip \cdot x} \langle0|T\left\{\rm{J}^{\mathrm{P_{cc}}}(x)\bar{\rm{J}}^{\mathrm{P_{cc}}}(0)\right\}|0\rangle _\gamma \, , \\
\Pi_{\mu\nu}(p,q)&=i\int d^4x e^{ip \cdot x} \langle0|T\left\{\rm{J}_\mu^{\mathrm{P_{cc}^\ast}}(x)\bar{\rm{J}}_\nu^{\mathrm{P_{cc}^\ast}}(0)\right\}|0\rangle _\gamma \,. \label{Pc101}
\end{align} 
Here, the interpolating currents ${\mathrm{J^{P_{cc}}}}(x)$ and $\mathrm{J_\mu^{P^\ast_{cc}}}(x)$ correspond to the spin-1/2 and spin-3/2 $D^{(\ast)} \Lambda_c^{(\ast)}$ pentaquark states, respectively. These currents serve as the fundamental building blocks for constructing the correlation functions within the QCD light-cone sum rule framework. The explicit forms of the relevant interpolating currents, which are essential for the subsequent analysis, are presented below~\cite{Duan:2024uuf}:
\begin{align}
\rm{J}^{ D \Lambda_c}(x)& =    \Big[\bar d^d(x)i \gamma_5 c^d(x)\Big]\Big[\varepsilon^{abc} u^{a^T}(x)C\gamma_\alpha c^b(x)  \gamma^\alpha\gamma_5 d^c(x)\Big]
 -  \Big[\bar d^d(x)i \gamma_5 c^d(x)\Big]  
 \Big[\varepsilon^{abc} d^{a^T}(x) C\gamma_\alpha c^b(x) 
 \gamma^\alpha\gamma_5 u^c(x)\Big] \, , \label{curpcs1}\\
 \nonumber\\
\rm{J}_{\mu}^{ D^\ast \Lambda_c}(x)& 
=\Big[\bar d^d(x)\gamma_\mu c^d(x)\Big]\Big[\varepsilon^{abc} u^{a^T}(x)C\gamma_\alpha c^b(x)  \gamma^\alpha\gamma_5 d^c(x)\Big]
 - \Big[\bar d^d(x)\gamma_\mu c^d(x)\Big]  
 \Big[\varepsilon^{abc} d^{a^T}(x) C\gamma_\alpha c^b(x) 
 \gamma^\alpha\gamma_5 u^c(x)\Big]\,, \\
 \nonumber\\
\rm{J}_{\mu}^{ D \Lambda_c^\ast}(x)&=  \Big[ \bar d^d(x)i \gamma_5 c^d(x)\Big]\Big[\varepsilon^{abc} u^{a^T}(x)C\gamma_\mu c^b(x)  d^c(x)\Big] - \Big[\bar d^d(x)i \gamma_5 c^d(x)\Big]  
 \Big[\varepsilon^{abc} u^{a^T}(x) C\gamma_\mu d^b(x) 
  c^c(x)\Big] \, . 
 \label{curpcs3}
  \end{align}

The interpolating currents for the $D^{(\ast)} \Lambda_c^{(\ast)}$ molecular pentaquarks are constructed as products of color-singlet charmed meson and baryon operators. These currents are specifically designed to have optimal overlap with the molecular configurations. Their construction as products of color-singlet operators, combined with the exact isospin $I=0$ symmetry, ensures strong preferential coupling to the molecular states rather than compact pentaquark configurations. The baryon components follow established forms: the $\Lambda_c$ current uses the standard spin-$\frac{1}{2}$ structure with $\gamma^\alpha\gamma_5$ projection, while the $\Lambda_c^\ast$ current employs a spin-$\frac{3}{2}$ configuration built from a vector diquark ($\gamma_\mu$) combined with a spectator quark, creating the necessary Rarita-Schwinger structure for a spin-$\frac{3}{2}$ baryon. The second terms in each baryon current ensure exact isospin $I=0$ symmetry through systematic $u \leftrightarrow d$ quark interchange. This precise isospin structure particularly enhances the coupling to molecular states. This preference arises because compact pentaquarks typically exhibit different color and spatial correlations, whereas our current construction explicitly mimics the meson-baryon molecular structure. The meson components $\bar d^d i\gamma_5 c^d$ (pseudoscalar) and $\bar d^d \gamma_\mu c^d$ (vector) represent the $D$ and $D^\ast$ mesons, respectively. These molecular combinations properly yield the expected $J^P$ quantum numbers: $\frac{1}{2}^-$ for $D\Lambda_c$, $\frac{3}{2}^-$ for $D^\ast\Lambda_c$, and $\frac{3}{2}^-$ for $D\Lambda_c^\ast$. While the currents may still have some coupling to other states with the same quantum numbers, the QCD sum rule methodology with continuum subtraction effectively isolates the contribution of the lowest-lying molecular states.

Within the framework of the QCD light-cone sum rule, the correlation function is initially expressed in terms of the physical parameters of the hadron. To connect it with hadronic properties, a complete set of intermediate states possessing the same quantum numbers as the interpolating currents, $\mathrm{J^{P_{cc}}}(x)$ and $\mathrm{J_\mu^{P^\ast_{cc}}}(x)$, is inserted into the correlation function. By employing the dispersion relation, the correlation function can then be represented as
 \begin{align}\label{edmn02}
\Pi^{Ph}(p,q)&=\frac{\langle0\mid \rm{J}^{\mathrm{P_{cc}}}(x) \mid
{\mathrm{P_{cc}}}(p, s) \rangle}{[p^{2}-m_{\mathrm{P_{cc}}}^{2}]}
\langle {\mathrm{P_{cc}}}(p, s)\mid
{\mathrm{P_{cc}}}(p+q, s)\rangle_\gamma 
\frac{\langle {\mathrm{P_{cc}}}(p+q, s)\mid
\bar{ \rm{J}}^{\mathrm{P_{cc}}}(0) \mid 0\rangle}{[(p+q)^{2}-m_{\mathrm{P_{cc}}}^{2}]}+ \cdots , \\
\Pi^{Ph}_{\mu\nu}(p,q)&=\frac{\langle0\mid  \rm{J}_{\mu}^{\mathrm{P_{cc}^\ast}}(x)\mid
{\mathrm{P_{cc}^\ast}}(p,s)\rangle}{[p^{2}-m_{{\mathrm{P_{cc}^\ast}}}^{2}]}
\langle {\mathrm{P_{cc}^\ast}}(p,s)\mid
{\mathrm{P_{cc}^\ast}}(p+q,s)\rangle_\gamma 
\frac{\langle {\mathrm{P_{cc}^\ast}}(p+q,s)\mid
\bar{\rm{J}}_{\nu}^{\mathrm{P_{cc}^\ast}}(0)\mid 0\rangle}{[(p+q)^{2}-m_{{\mathrm{P_{cc}^\ast}}}^{2}]}+ \cdots .\label{Pc103}
\end{align}

For the subsequent calculations, it is essential to evaluate the matrix elements defined in Eqs.~(\ref{edmn02}) and (\ref{Pc103}). These elements can be expressed in terms of the corresponding hadronic observables as follows:
%
\begin{align}
\label{edmn005}
\langle0\mid \rm{J}^{\mathrm{P_{cc}}}(x)\mid {\mathrm{P_{cc}}}(p, s)\rangle=&\lambda_{\mathrm{P_{cc}}} \gamma_5 \, \nu(p,s),\\
\langle0\mid \rm{J}_{\mu}^{\mathrm{P_{cc}^\ast}}(x)\mid {\mathrm{P_{cc}^\ast}}(p,s)\rangle&=\lambda_{{\mathrm{P_{cc}^\ast}}}u_{\mu}(p,s),\\
\langle {\mathrm{P_{cc}}}(p, s)\mid {\mathrm{P_{cc}}}(p+q, s)\rangle_\gamma &=\varepsilon^\alpha \,\bar \nu(p, s)\bigg[\big[f_1(q^2)+f_2(q^2)\big] \gamma_\alpha +f_2(q^2)\frac{(2p+q)_\alpha}{2 m_{\mathrm{P_{cc}}}}\bigg]\,\nu(p+q, s), \\
\langle {\mathrm{P_{cc}^\ast}}(p,s)\mid {\mathrm{P_{cc}^\ast}}(p+q,s)\rangle_\gamma &=-e \, \bar u_{\alpha}(p,s)\bigg[F_{1}(q^2)g_{\alpha\beta}\eslash 
-
\frac{1}{2m_{{\mathrm{P_{cc}^\ast}}}} 
\Big[F_{2}(q^2)g_{\alpha\beta} \eslash\qslash
+F_{4}(q^2)\frac{q_{\alpha}q_{\beta}\eslash\qslash}{(2m_{{\mathrm{P_{cc}^\ast}}})^2}\Big]
\nonumber\\
&+
F_{3}(q^2)\frac{1}{(2m_{{\mathrm{P_{cc}^\ast}}})^2}q_{\alpha}q_{\beta}\eslash \bigg] 
u_{\beta}(p+q,s).
\label{matelpar}
\end{align}
Here, $\nu(p,s)$ and $u_{\mu(\beta)}(p,s)$ represent the Dirac and Rarita-Schwinger spinors for the $\mathrm{P_{cc}}$ and $\mathrm{P_{cc}^\ast}$ states, respectively. The parameters $\lambda_{\mathrm{P_{cc}}}$ and $\lambda_{\mathrm{P_{cc}^\ast}}$ denote the pole residues, while $f_i(q^2)$ and $F_i(q^2)$ are the electromagnetic form factors characterizing the structure of the $\mathrm{P_{cc}}$ and $\mathrm{P_{cc}^\ast}$ transitions, in order.  For detailed derivations of the matrix elements $\langle \mathrm{P_{cc}}(p, s) \mid \mathrm{P_{cc}}(p+q, s) \rangle_\gamma$ and $\langle \mathrm{P_{cc}^\ast}(p, s) \mid \mathrm{P_{cc}^\ast}(p+q, s) \rangle_\gamma$, we refer the reader to Refs.~\cite{Leinweber:1990dv, Weber:1978dh, Nozawa:1990gt, Pascalutsa:2006up, Ramalho:2009vc}.

By employing the relations introduced above, one can construct the phenomenological representation of the correlation functions relevant to the magnetic dipole moments. This involves expressing the correlation functions in terms of hadronic parameters, including the spinors of the $\mathrm{P_{cc}}$ and $\mathrm{P_{cc}^\ast}$ states, their pole residues, and the associated form factors. Subsequently, standard mathematical manipulations, such as isolating specific Lorentz structures and applying the dispersion relation, allow the separation of contributions from different spin components and transitions. Through these steps, the phenomenological side of the correlation functions can be systematically expressed in a form suitable for matching with the QCD side and for extracting the magnetic dipole moments of the pentaquark states.
\begin{align}
\label{edmn050}
\Pi^{Ph}(p,q)&=\frac{\lambda^2_{\mathrm{P_{cc}}}}{[(p+q)^2-m^2_{\mathrm{P_{cc}}}][p^2-m^2_{\mathrm{P_{cc}}}]}
  \bigg[\Big(f_1(q^2)+f_2(q^2)\Big)\Big(
  2 (\varepsilon . p) \pslash -
  m_{\mathrm{P_{cc}}}\,\eslash \pslash
  -m_{\mathrm{P_{cc}}}\,\eslash \qslash
  +\pslash\eslash\qslash
  \Big)+ \cdots \bigg],\\
\Pi^{Ph}_{\mu\nu}(p,q)&=\frac{\lambda_{_{{\mathrm{P_{cc}^\ast}}}}^{2}}{[(p+q)^{2}-m_{_{{\mathrm{P_{cc}^\ast}}}}^{2}][p^{2}-m_{_{{\mathrm{P_{cc}^\ast}}}}^{2}]} 
\bigg[ F_{1}(q^2) \, g_{\mu\nu}\pslash\eslash\qslash 
- F_{2}(q^2) \, m_{{\mathrm{P_{cc}^\ast}}}g_{\mu\nu}\eslash\qslash
+ F_{3}(q^2)\,
\frac{1}{2m_{{\mathrm{P_{cc}^\ast}}}}q_{\mu}q_{\nu}\eslash\qslash \nonumber\\
&
+ F_{4}(q^2)\,
\frac{1}{4m_{{\mathrm{P_{cc}^\ast}}}^3}(\varepsilon.p)q_{\mu}q_{\nu}\pslash\qslash 
+
\cdots 
\bigg]. \label{final phenpart1}
\end{align}

In order to calculate the magnetic form factors $F_M(q^2)$ and $G_M(q^2)$ for the $D^{(\ast)} \Lambda_c^{(\ast)}$ pentaquarks, corresponding to the $\mathrm{P_{cc}}$ and $\mathrm{P_{cc}^\ast}$ states, respectively, it is necessary to express them in terms of the previously defined transition form factors, $f_i(q^2)$ and $F_i(q^2)$. The explicit relations between these sets of form factors are presented below \cite{Leinweber:1990dv, Weber:1978dh, Nozawa:1990gt, Pascalutsa:2006up, Ramalho:2009vc}:
\begin{align}
\label{edmn07}
F_M(q^2) &= f_1(q^2) + f_2(q^2),\\
G_{M}(q^2) &= \left[ F_1(q^2) + F_2(q^2)\right] ( 1+ \frac{4}{5}
\eta ) -\frac{2}{5} \left[ F_3(q^2)  
+
F_4(q^2)\right] \eta \left( 1 + \eta \right),
\end{align}  
where $\eta = -\frac{q^2}{4m^2_{{\mathrm{P_{cc}^\ast}}}}$.    
The above relations allow for the determination of the electromagnetic form factors of the $ D^{(\ast)} \Lambda_c^{(\ast)}$. Since our analysis involves a real photon ($q^2 = 0$), it is necessary to consider the corresponding form factors relevant to the magnetic dipole moment. The explicit expressions for these form factors are provided as follows:
\begin{align}
\label{edmn08}
\mu_{\mathrm{P_{cc}}} &= \frac{ e}{2\, m_{\mathrm{P_{cc}}}} \,[f_1(0)+f_2(0)],~~~~~%
\\
\mu_{{\mathrm{P_{cc}^\ast}}}&=\frac{e}{2m_{{\mathrm{P_{cc}^\ast}}}}[F_1(0)+F_2(0)].
\end{align}

On the theoretical side, the quark fields are first contracted using Wick's theorem, followed by the application of the operator product expansion (OPE). After performing the contractions of the quark field operators, the correlation function can be expressed in terms of the propagators of both light and heavy quarks, as well as the photon distribution amplitudes (DAs). The outcome of these theoretical manipulations is then obtained as follows:
\begin{align}
\label{QCD1}
\Pi^{\rm{Th}-\mathrm{ D \Lambda_c}}(p,q)&= -i \,\varepsilon^{abc} \varepsilon^{a^{\prime}b^{\prime}c^{\prime}}\, \int d^4x \, e^{ip\cdot x} 
 \langle 0\mid \Big\{ \mbox{Tr}\Big[\gamma_5  S_{c}^{dd^\prime}(x) \gamma_5 S_{d}^{d^\prime d}(-x) \Big]  
\mbox{Tr}\Big[\gamma_{\alpha} S_c^{bb^\prime}(x) \gamma_{\beta}  
  \widetilde S_{u}^{aa^\prime}(x)\Big] 
 \nonumber\\
& \times ( \gamma_5\gamma^{\alpha} S_{d}^{cc^\prime}(x) \gamma^{\beta} \gamma_5 ) + \mbox{Tr}\Big[\gamma_5  S_{c}^{dd^\prime}(x) \gamma_5 S_{d}^{d^\prime d}(-x) \Big]  
\mbox{Tr}\Big[\gamma_{\alpha} S_c^{bb^\prime}(x) \gamma_{\beta}  
  \widetilde S_{d}^{aa^\prime}(x)\Big] 
  \nonumber\\
& \times ( \gamma_5\gamma^{\alpha} S_{u}^{cc^\prime}(x) \gamma^{\beta} \gamma_5 )\Big\}
 \mid 0 \rangle _\gamma \,,\\
 \nonumber\\
%
\Pi_{\mu\nu}^{\rm{Th}-\mathrm{ D^\ast \Lambda_c}}(p,q)&= -i \,\varepsilon^{abc} \varepsilon^{a^{\prime}b^{\prime}c^{\prime}}\, \int d^4x \, e^{ip\cdot x} 
 \langle 0\mid \Big\{ \mbox{Tr}\Big[\gamma_{\alpha} S_{c}^{bb^\prime}(x) \gamma_{\beta} \widetilde S_{u}^{aa^\prime }(x)\Big]  
\mbox{Tr}\Big[\gamma_{\mu} S_c^{dd^\prime}(x) \gamma_{\nu}  
   S_{d}^{d^\prime d}(-x)\Big] 
 \nonumber\\
& \times (\gamma_5 \gamma^{\alpha} S_{d}^{cc^\prime}(x)   \gamma^{\beta} \gamma_5)       
+\mbox{Tr}\Big[\gamma_{\alpha} S_{c}^{bb^\prime}(x) \gamma_{\beta} \widetilde S_{d}^{aa^\prime }(x)\Big]   
\mbox{Tr}\Big[\gamma_{\mu} S_c^{dd^\prime}(x) \gamma_{\nu}  
   S_{d}^{d^\prime d}(-x)\Big]
   \nonumber\\
& \times (\gamma_5 \gamma^{\alpha} S_{u}^{cc^\prime}(x)   \gamma^{\beta} \gamma_5)
 \Big\} 
\mid 0 \rangle _\gamma \,, \label{QCD2}
\\
\nonumber\\
%
\Pi_{\mu\nu}^{\rm{Th}-\mathrm{ D \Lambda_c^\ast}}(p,q)&=i\,\varepsilon^{abc} \varepsilon^{a^{\prime}b^{\prime}c^{\prime}}\, \int d^4x \, e^{ip\cdot x} 
 \langle 0\mid \Big\{ \mbox{Tr}\Big[\gamma_5  S_{c}^{dd^\prime}(x)\gamma_5 S_{d}^{d^\prime d}(-x) \Big]  
\mbox{Tr}\Big[\gamma_{\mu} S_c^{bb^\prime}(x) \gamma_{\nu}  
  \widetilde S_{u}^{aa^\prime}(x)\Big]  S_{d}^{cc^\prime}(x)
 \nonumber\\
&
  + \mbox{Tr}\Big[\gamma_5  S_{c}^{dd^\prime}(x)\gamma_5 S_{d}^{d^\prime d}(-x) \Big]  
\mbox{Tr}\Big[\gamma_{\mu} S_d^{bb^\prime}(x) \gamma_{\nu}  
  \widetilde S_{u}^{aa^\prime}(x)\Big]  S_{c}^{cc^\prime}(x)
 \Big\} 
\mid 0 \rangle _\gamma \,,\label{QCD3}
\end{align}
where   
$\widetilde{S}_{c(q)}^{ij}(x)=CS_{c(q)}^{ij\mathrm{T}}(x)C$. The light ($S_{q}(x)$) and charm ($S_{c}(x)$) quark propagators relevant for this analysis are expressed as follows~\cite{Balitsky:1987bk, Belyaev:1985wza}:
\begin{align}
\label{edmn13}
S_{q}(x)&= \frac{1}{2 \pi x^2}\Big(i \frac{\xslash}{x^2}- \frac{m_q}{2}\Big) 
-i\frac { g_s }{16 \pi^2 x^2} \int_0^1 du \, G^{\mu \nu} (ux)
\bigg[\bar u \rlap/{x} 
\sigma_{\mu \nu} + u \sigma_{\mu \nu} \rlap/{x}
 \bigg],\\
%
S_{c}(x)&=\frac{m_{c}^{2}}{4 \pi^{2}} \bigg[ \frac{K_{1}\big(m_{c}\sqrt{-x^{2}}\big) }{\sqrt{-x^{2}}}
+i\frac{{\xslash}~K_{2}\big( m_{c}\sqrt{-x^{2}}\big)}
{(\sqrt{-x^{2}})^{2}}\bigg]
-i\frac{m_{c}\,g_{s} }{16\pi ^{2}}  \int_0^1 du \,G^{\mu \nu}(ux)\bigg[ (\sigma _{\mu \nu }{\xslash}
+{\xslash}\sigma _{\mu \nu }) 
    \frac{K_{1}\big( m_{c}\sqrt{-x^{2}}\big) }{\sqrt{-x^{2}}}
   \nonumber\\
  &
 +2\sigma_{\mu \nu }K_{0}\big( m_{c}\sqrt{-x^{2}}\big)\bigg],
 \label{edmn14}
\end{align}%
where $G^{\mu\nu}(x)$ is the gluon field-strength tensor, and $K_n$'s being the modified second type Bessel functions.  

On the theoretical side, the correlation function receives two distinct types of contributions. The first corresponds to the perturbative contribution, arising from photon emission at short distances, while the second represents the non-perturbative contribution, associated with photon emission at long distances. In order to isolate the short-distance contributions, the following expression is employed: 
\begin{align}
\label{free}
S^{free}(x) \rightarrow \int d^4z\, S^{free} (x-z)\,\rlap/{\!A}(z)\, S^{free} (z)\,,
\end{align}
where $S^{free}(x)$ denotes the first term of the light and heavy quark propagators, and the four propagators that survive in Eqs.~(\ref{QCD1})-(\ref{QCD3}) are considered to be free. 
To account for the long-distance contributions, the following approach should be applied:

 \begin{align}
\label{edmn21}
S_{\alpha\beta}^{ab}(x) \rightarrow -\frac{1}{4} \Big[\bar{q}^a(x) \Gamma_i q^b(0)\Big]\Big(\Gamma_i\Big)_{\alpha\beta},
\end{align}
where $\Gamma_i$  stands for the full set of gamma matrices, $\Gamma_i$ $=$ $\{\mathbb 1$, $\gamma_5$, $\gamma_\mu$, $i\gamma_5 \gamma_\mu$, $\sigma_{\mu\nu}/2\}$, and the four propagators that remain in Eqs.~(\ref{QCD1})-(\ref{QCD3}) are considered to be full propagators. 
Following the procedures outlined in Eq.~(\ref{edmn21}), new terms such as $\langle \gamma(q) | \bar{q}(x) \Gamma_i G_{\alpha\beta} q(0) | 0 \rangle$ and $\langle \gamma(q) | \bar{q}(x) \Gamma_i q(0) | 0 \rangle$ arise as essential elements for the continuation of the analysis. These matrix elements, defined in terms of the photon DAs, play a crucial role in evaluating the long-distance contributions (for further details, see Ref.~\cite{Ball:2002ps}). The theoretical side of the correlation functions is calculated using the expressions in Eqs.~(\ref{QCD1})–(\ref{edmn21}) and subsequently transformed into momentum space via the Fourier transform. By systematically applying these intricate and technical steps, the theoretical expressions for the magnetic dipole moments are finally obtained.

The QCD light-cone sum rules for the magnetic dipole moments can then be established based on the principle of quark-hadron duality, which requires that the correlation functions calculated on the phenomenological and theoretical sides coincide. At this stage, an additional step is necessary to enhance the contribution of the ground state while suppressing the effects of the continuum and higher excited states. To achieve this, the standard procedures of continuum subtraction and Borel transformation are applied within the QCD light-cone sum rules framework. The resulting expressions for the magnetic dipole moments, obtained after performing these manipulations, are presented below:

\begin{align}
\label{edmn15}
&\mu_{ D \Lambda_c} \,\lambda^2_{ D \Lambda_c}=e^{\frac{m^2_{ D \Lambda_c}}{\rm{M^2}}}\, \rho_1 (\rm{M^2},\rm{s_0}),\\
&\mu_{ D \Lambda_c^\ast} \,\lambda^2_{ D \Lambda_c^\ast}  =e^{\frac{m^2_{ D \Lambda_c^\ast}}{\rm{M^2}}}\, \rho_2 (\rm{M^2},\rm{s_0}),\\
&\mu_{ D^\ast \Lambda_c} \,\lambda^2_{ D^\ast \Lambda_c} =e^{\frac{m^2_{ D^\ast \Lambda_c}}{\rm{M^2}}}\, \rho_3 (\rm{M^2},\rm{s_0}). \label{edmn015}
\end{align}

As an illustrative example, the explicit form of the function $\rho_1(\rm{M}^2,\rm{s}_0)$ is given in Eq.~(\ref{sumrules}). 
This function encodes the contributions of different quark and gluon configurations to the sum rule and serves 
as a key ingredient in the determination of the magnetic dipole moments. Although the explicit forms of 
$\rho_2(\rm{M}^2,\rm{s}_0)$, $\rho_3(\rm{M}^2,\rm{s}_0)$, and other related functions are not shown here, 
they follow analogous structures and are derived using the same procedures. These functions are obtained after 
applying the Borel transformation and continuum subtraction, and they incorporate both perturbative and 
non-perturbative contributions from the correlation function.

 \begin{align}
\label{sumrules}
 \rho_1 (\rm{M^2},\rm{s_0}) &= \frac{61}{2 ^{27}\times 5^3\times 7^2 \pi^7} \Big[ \big(4 e_c - 9 (e_u + e_d)\big) I_{07}\Big] \nonumber\\
 &+\frac {m_c\langle g_s^2 G^2\rangle \langle \bar q q \rangle } {2^{21} \times 3^4 \pi^3} (e_u+e_d) \Big[\big(20 \mathbb A(u_0)  + 78 I_3(\mathcal S)  - 90 I_3(\mathcal{\tilde S})\big) I_{03} \Big]\nonumber\\
        & - \frac {\langle g_s^2 G^2\rangle f_ {3 \gamma} } {2^{23}\times 3^3 \times 5 \pi^5} (e_u+e_d) \Big [ \big(13 I_1(\mathcal V) + 10 \psi^a(u_0)\big)I_{04}\Big]\nonumber\\
   &-\frac {17 m_c \langle g_s^2 G^2\rangle \langle \bar q q \rangle \chi} {2^{21} \times 3^4 \pi^3} (e_u+e_d) \Big[ 
  I_{04} \varphi_\gamma(u_0) \Big]\nonumber\\
   & + \frac { m_c \langle \bar q q \rangle } {2^{24} \times 5^2 \pi^3} (e_u+e_d) \Big[ \big(7 \mathbb A(u_ 0)  + 28 I_ 3(\mathcal S)  - 
 48 I_ 3(\mathcal{\tilde S})\big) I_{05}  \Big]\nonumber\\
      & - \frac { f_ {3\gamma } } {2 ^{25}\times 3 \times 5^2 \times 7 \pi^5} (e_u+e_d)
   \Big [\big(140 I_1(\mathcal V) - 129 \psi^a(u_0)\big)I_{06}\Big] \nonumber\\
   & - \frac { m_c \langle \bar q q \rangle \chi} {2^{23} \times 5^2 \pi^3} (e_u+e_d) \Big[    I_{06} \varphi_\gamma(u_0) \Big],
\end{align}
where functions $I_{nm}$ and $I_i(\mathcal{F})$ are defined through the following expressions:

\begin{align}
 I_{nm}&= \int_{4 m_c^2}^{\rm{s_0}} ds \, s^n (s-4 m_c^2)^m\,e^{-s/\rm{M^2}},~
 \nonumber\\
 I_1(\mathcal{F})&=\int D_{\alpha_i} \int_0^1 dv~ \mathcal{F}(\alpha_{\bar q},\alpha_q,\alpha_g) \delta'(\alpha_ q +\bar v \alpha_g-u_0),\nonumber\\
    I_3(\mathcal{F})&=\int D_{\alpha_i} \int_0^1 dv~ \mathcal{F}(\alpha_{\bar q},\alpha_q,\alpha_g) \delta(\alpha_ q +\bar v \alpha_g-u_0),
 \end{align}
 with $\mathcal{F}$ denotes the relevant photon DAs. 

\end{widetext}

\section{Magnetic properties: Numerical results and Discussion}\label{numerical}

In this section, we present the numerical analysis of the expressions
derived for the magnetic dipole moments of $D^{(\ast)} \Lambda_c^{(\ast)}$ pentaquarks. 
For this purpose, a set of input parameters is required, which are listed 
in Table~\ref{inputparameter}. These include the charm-quark mass ($m_c$), 
the quark condensate ($\langle \bar q q \rangle$), the gluon condensate 
($\langle g_s^2 G^2 \rangle$), and the magnetic susceptibility of the quark 
condensate ($\chi$). In addition, the mass and pole residue of the states 
under investigation are employed. For the charm-quark mass, the value 
reported by the Particle Data Group (PDG) is adopted. The up- and down-quark 
masses ($m_u$ and $m_d$) are also taken into account, although their effects 
are negligible, and are therefore set to zero in the numerical evaluations. 
The values of the quark and  gluon condensates are chosen in line 
with previous QCD sum rule analyses. For the magnetic susceptibility of 
the quark condensate, we make use of the estimate obtained via the QCD 
light-cone sum rule approach. Since no experimental determinations exist 
for the masses and residues of the $D^{(\ast)} \Lambda_c^{(\ast)}$ pentaquarks, 
we employ the predictions of QCD sum rules, which are known to provide 
consistent and reliable estimates in the absence of experimental input. 
Finally, the photon DAs, which also enter the 
sum rules in Eq.~(\ref{sumrules}), are taken from Ref.~\cite{Ball:2002ps}.


 \begin{table}[htb]
	\addtolength{\tabcolsep}{10pt}
	\caption{Input parameters employed in the numerical calculations.}
	\label{inputparameter}
\begin{tabular}{l|ccc}
               \hline\hline
                \\
Parameters & Values \\
 \\
                                        \hline\hline
$m_c$&$ 1.27 \pm 0.02$ GeV  \cite{Workman:2022ynf}
                        \\
$m_{ D \Lambda_c}$&$  4.13^{+0.10}_{-0.09}$~GeV  \cite{Duan:2024uuf}
                                                \\
$m_{ D \Lambda_c^*}$&$ 4.31^{+0.11}_{-0.10}$~GeV \cite{Duan:2024uuf}
                                               \\
$m_{ D^* \Lambda_c} $&$  4.42^{+0.13}_{-0.12}$~GeV \cite{Duan:2024uuf}
                                              \\
$\lambda_{ D \Lambda_c}  $&$ (0.77^{+0.16}_{-0.16})\times 10^{-3} $\,\,GeV$^6$ \cite{Duan:2024uuf}
                       \\
$ \lambda_{ D \Lambda_c^*}   $&$ (0.95^{+0.21}_{-0.21})\times 10^{-3} $\,\,GeV$^6$ \cite{Duan:2024uuf}
                       \\
$\lambda_{ D^* \Lambda_c}   $&$(0.79^{+0.16}_{-0.15})\times 10^{-3} $\,\,GeV$^6$ \cite{Duan:2024uuf}
\\
$f_{3\gamma} $&$ -0.0039$\,\,GeV$^2$ \cite{Ball:2002ps}
                       \\
$\langle \bar qq\rangle $&$ (-0.24 \pm 0.01)^3 $\,\,GeV$^3$ \cite{Ioffe:2005ym}
                       \\
$ \langle g_s^2G^2\rangle  $&$ 0.48 \pm 0.14 $\,\,GeV$^4$ \cite{Narison:2018nbv}
                       \\
$\chi $&$ -2.85 \pm 0.5 $\,\,\,\,\,GeV$^{-2}$\cite{Rohrwild:2007yt}
                       \\
                                    \hline\hline
 \end{tabular}
\end{table}

\begin{widetext}

\begin{figure}[htb!]
\includegraphics[width=0.32\textwidth]{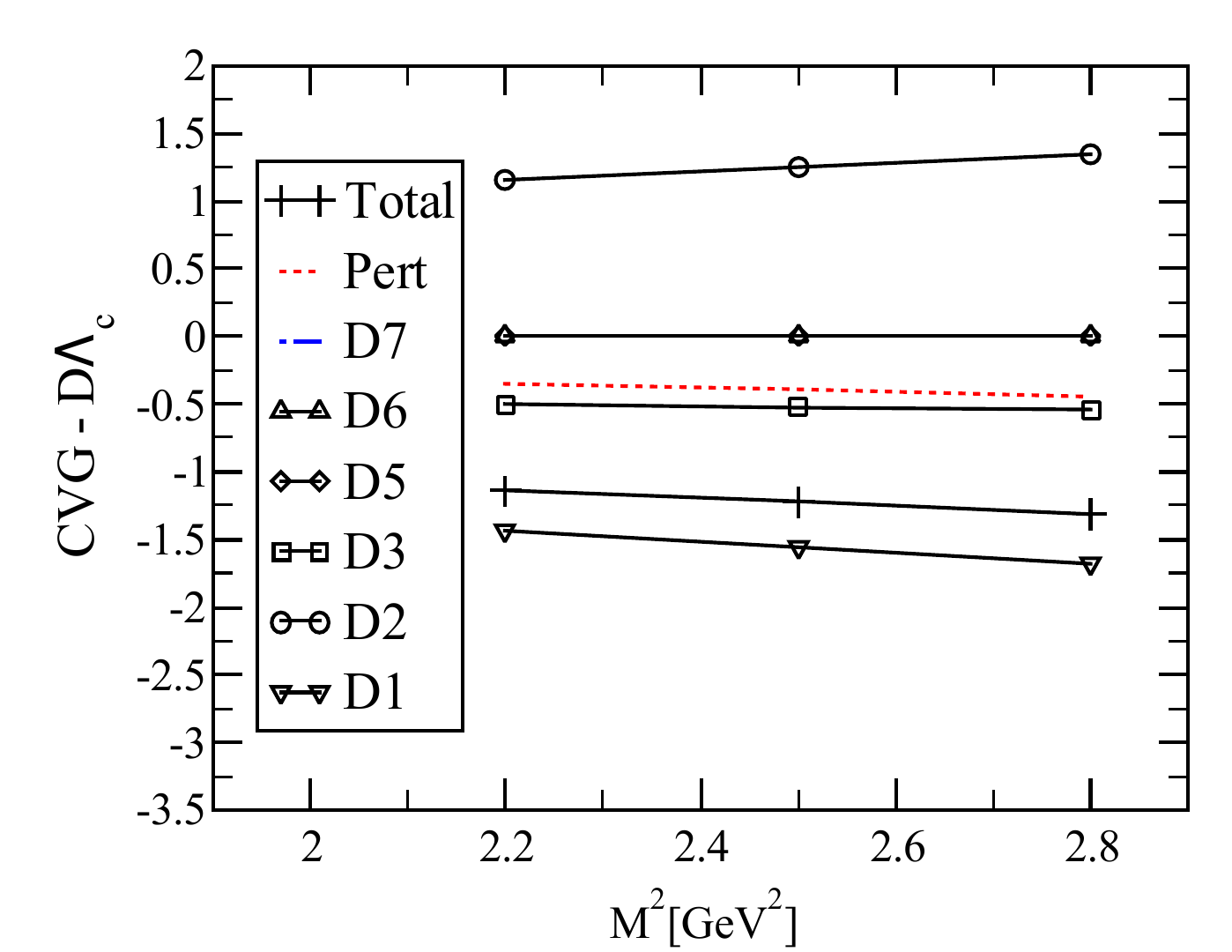} ~
\includegraphics[width=0.32\textwidth]{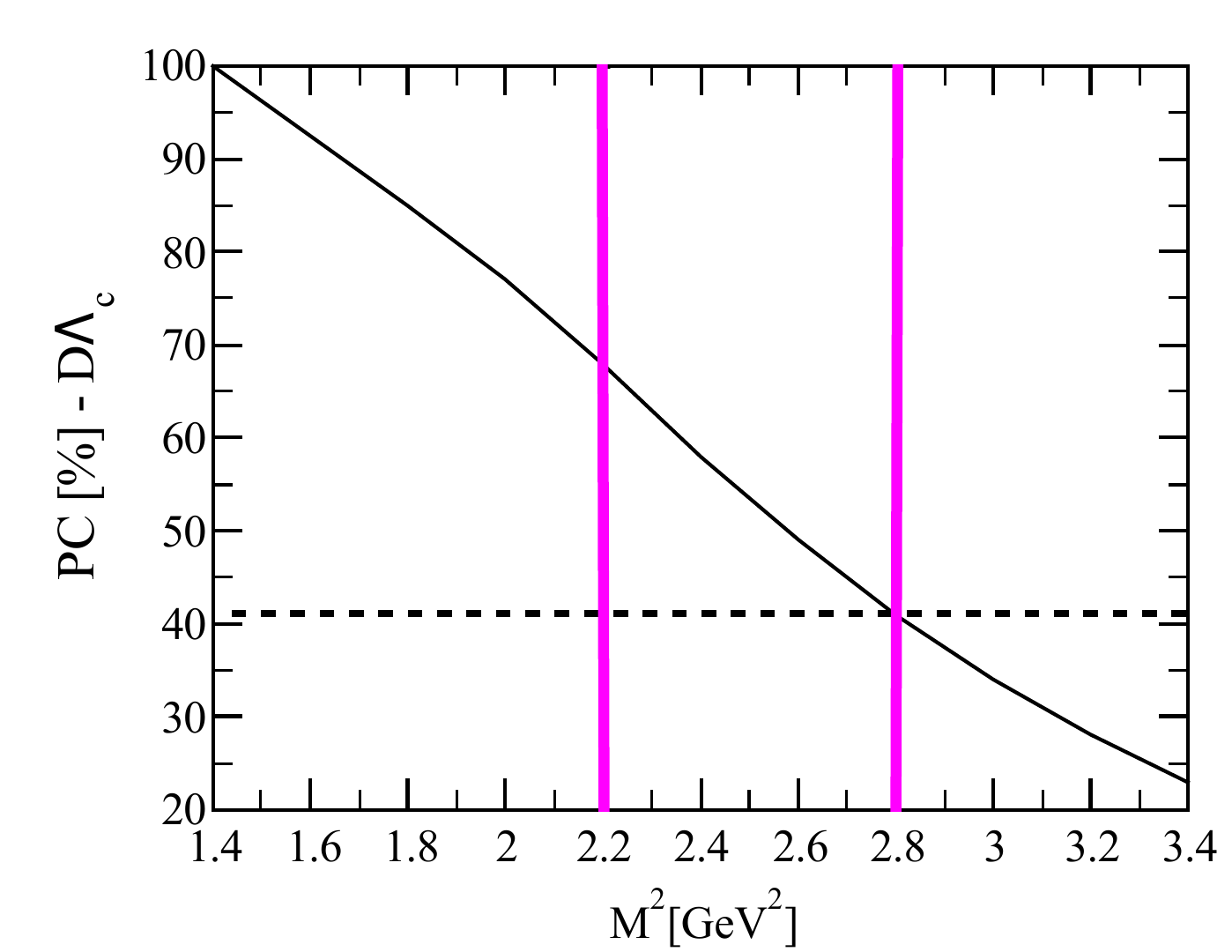} ~
\includegraphics[width=0.32\textwidth]{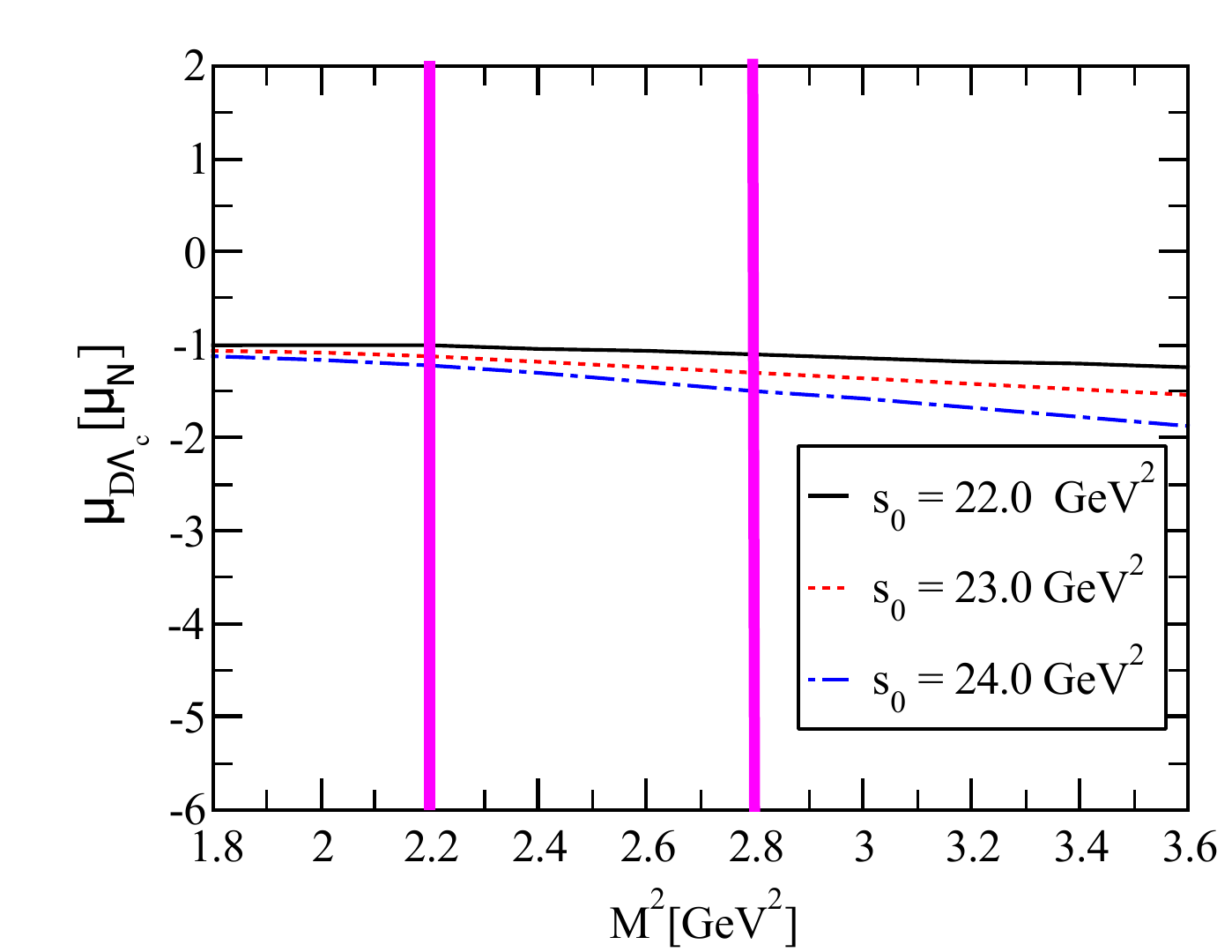}\\
\includegraphics[width=0.32\textwidth]{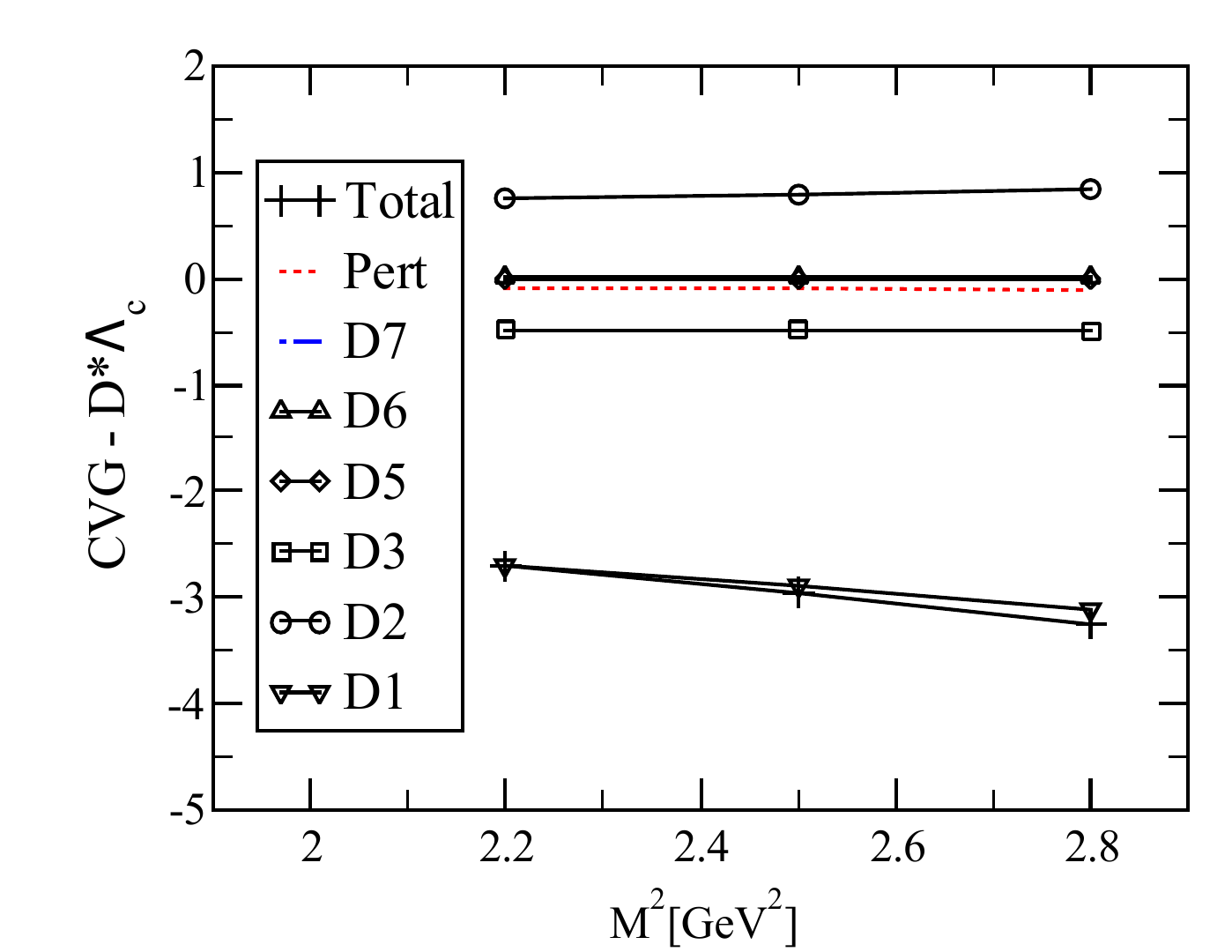}~
\includegraphics[width=0.32\textwidth]{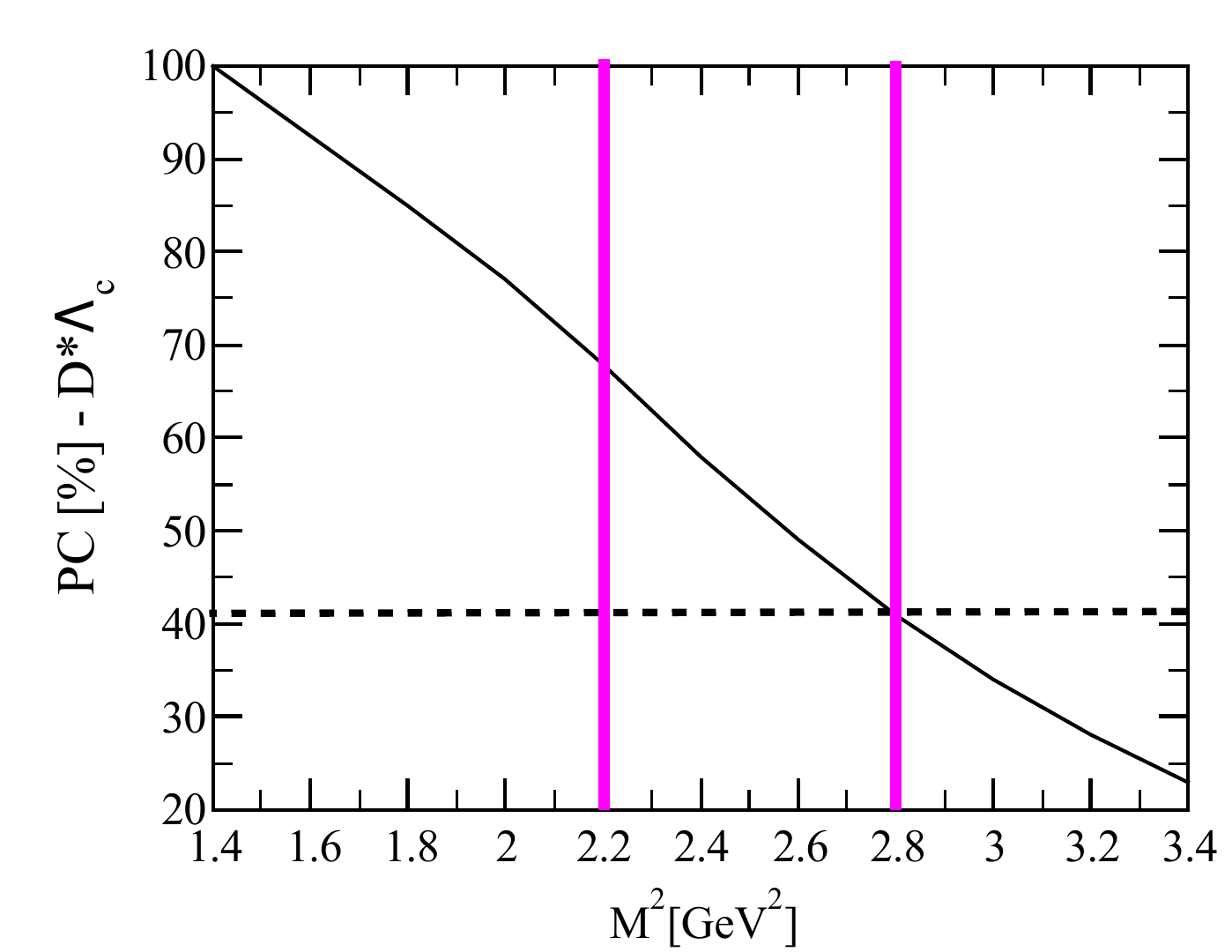} ~
\includegraphics[width=0.32\textwidth]{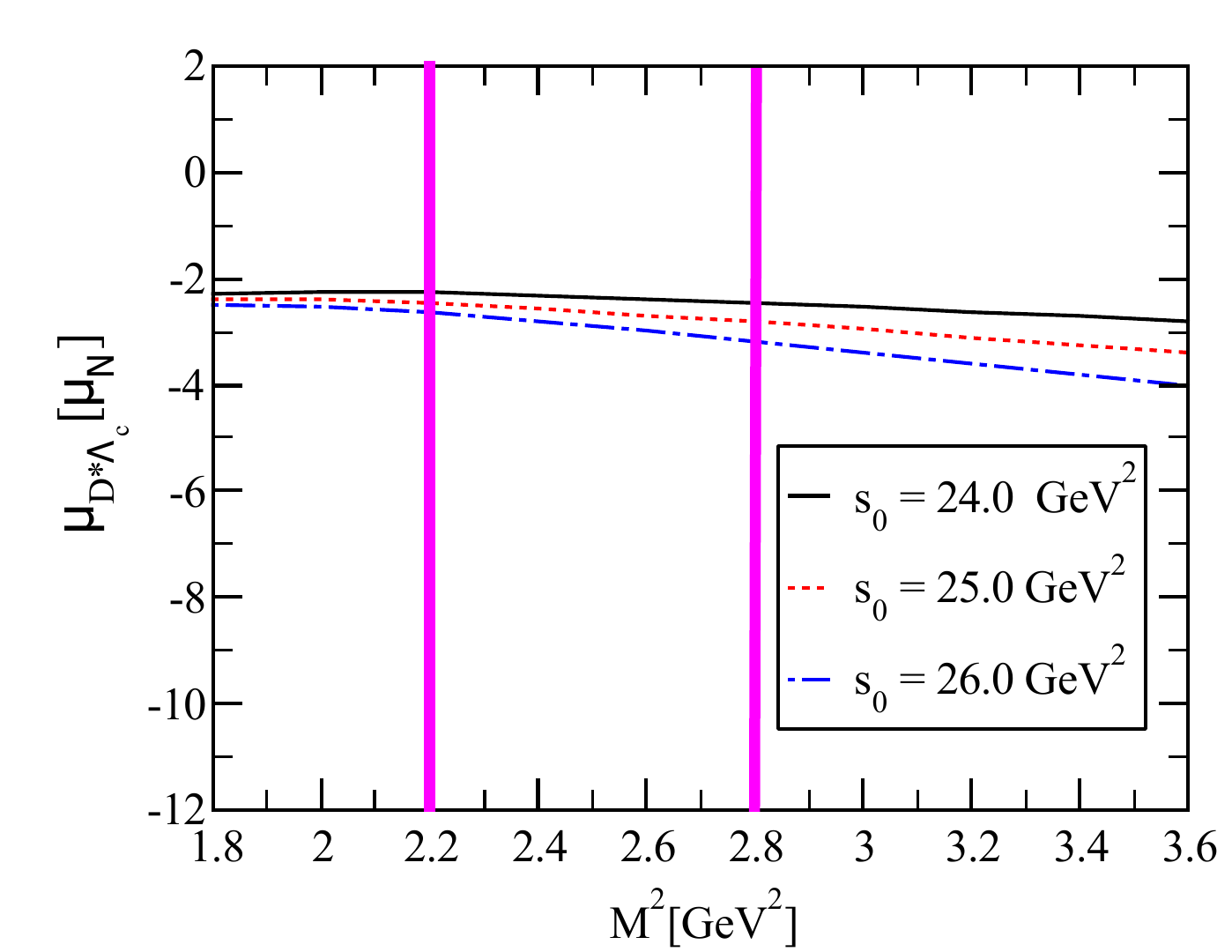}\\
\includegraphics[width=0.32\textwidth]{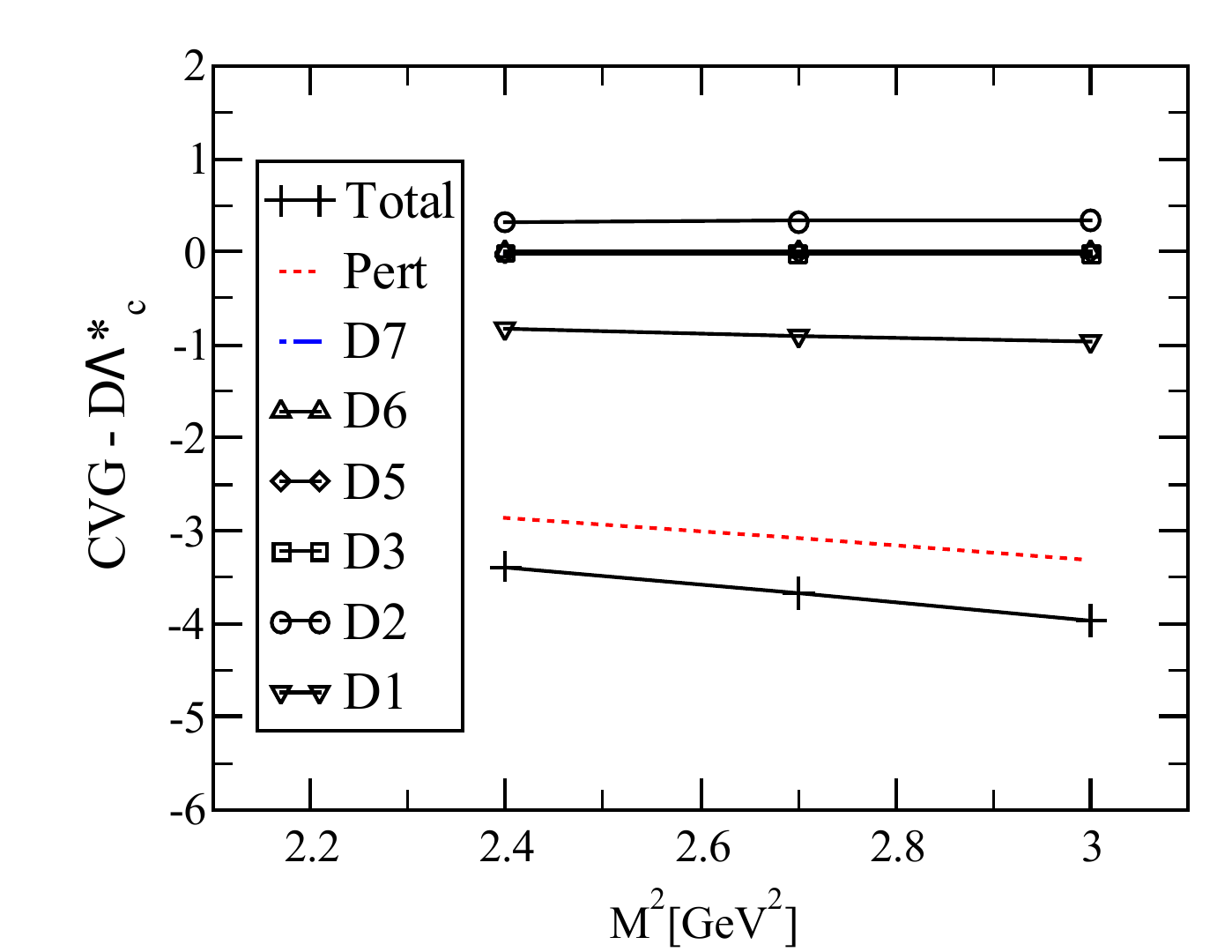}~
\includegraphics[width=0.32\textwidth]{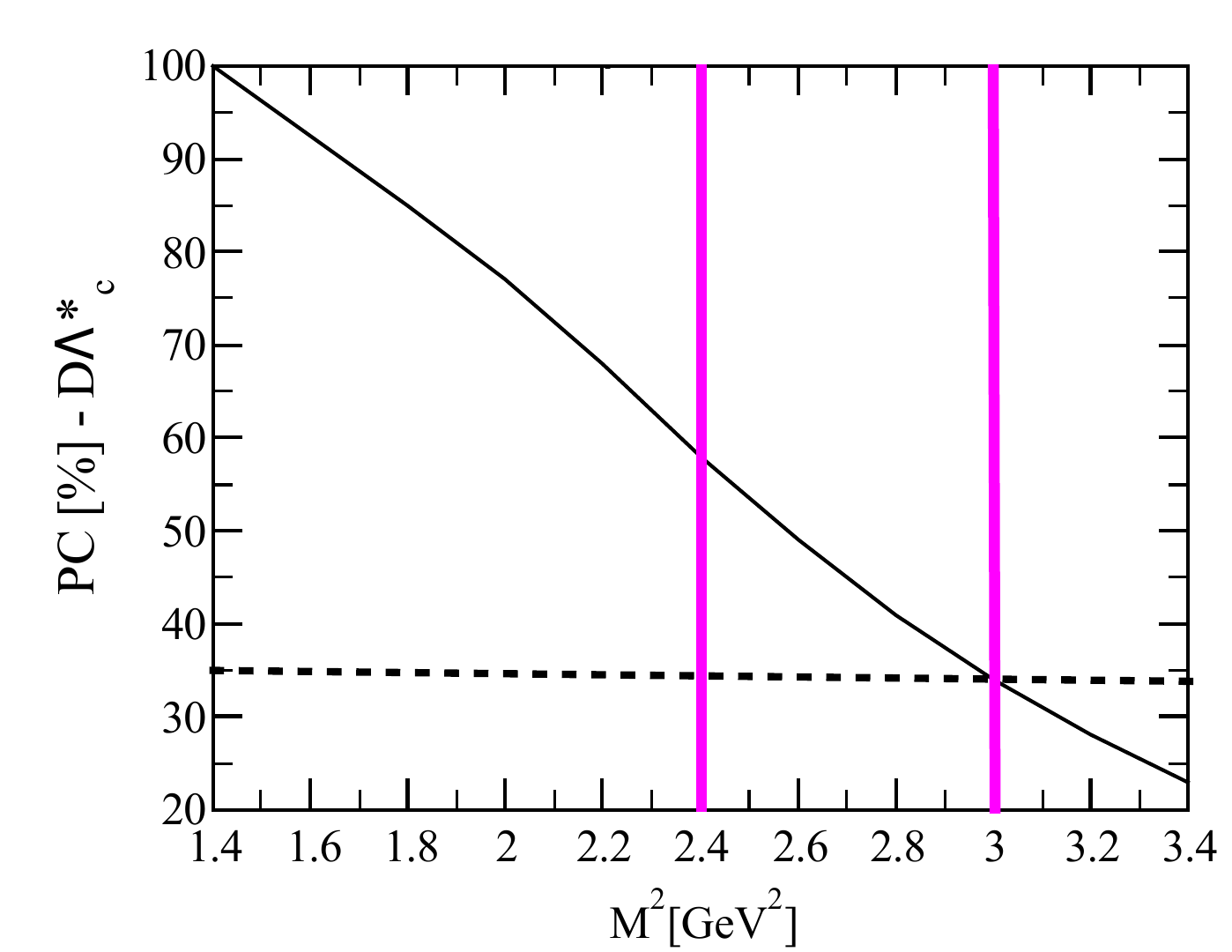} ~
\includegraphics[width=0.32\textwidth]{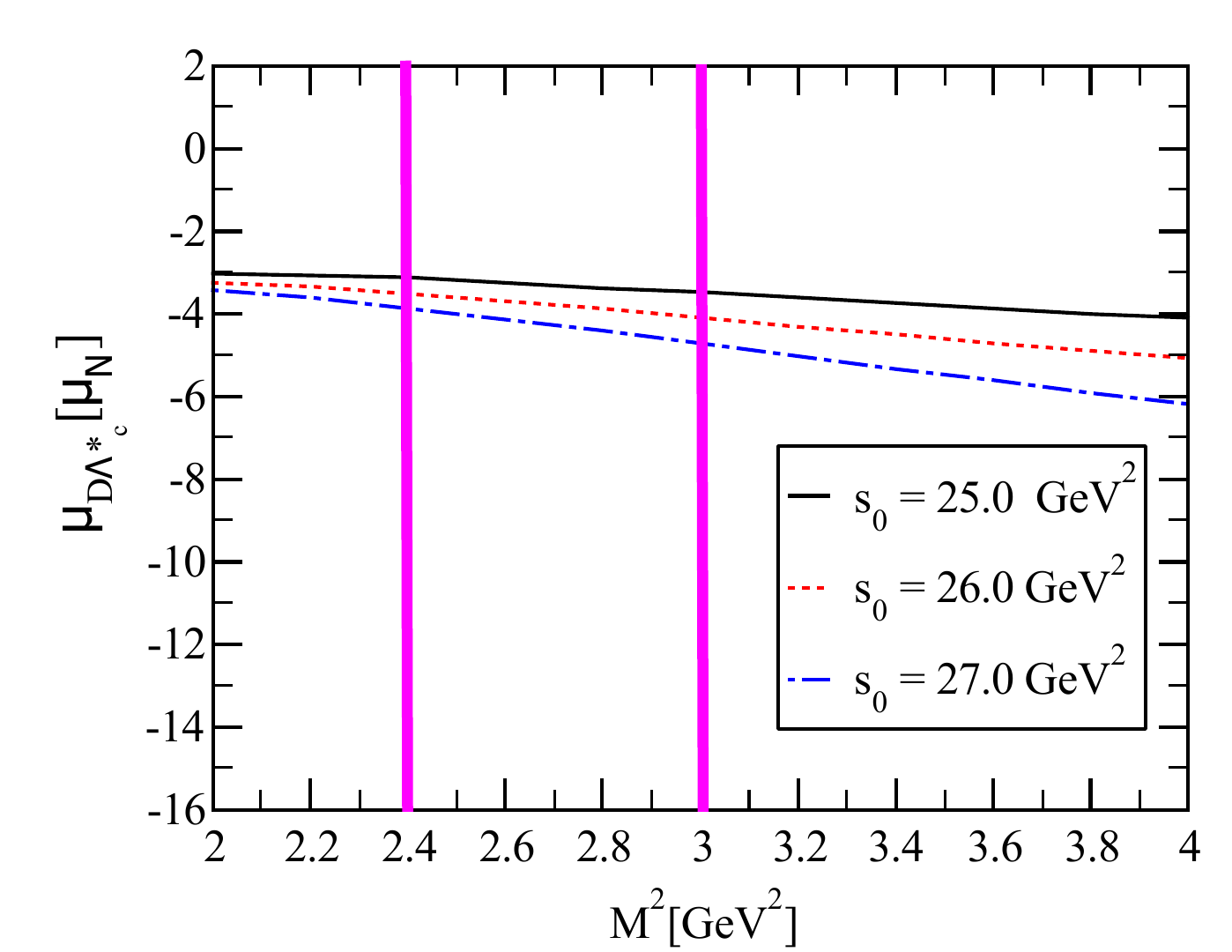}
 \caption{CVG analysis (left panels), PC (middle panels), and total value (right panels) for the magnetic dipole moments of the $D^{(\ast)} \Lambda_c^{(\ast)}$ pentaquarks versus $\rm{M^2}$ at fixed $\rm{s_0}$ values. The adopted Borel window is illustrated by the vertical lines in the middle and right panels, whereas the horizontal line in the middle panel marks the minimum PC value obtained within this region.}
 \label{Msqfig1}
  \end{figure}

  \end{widetext}

In addition to the parameters discussed above, two auxiliary inputs are
required: the Borel mass parameter $\rm{M^2}$ and the continuum threshold $\rm{s_0}$. The numerical analysis is carried out within the standard constraints of the QCD sum rule method, which guarantees the reliability
of the extracted results. The admissible range of $\rm{M^2}$ is fixed by
imposing two well-established criteria: pole dominance (PC) and the
convergence of the operator product expansion (CVG). These conditions can
be formulated as follows:
\begin{align}
 \mbox{PC} &=\frac{\rho_i (\rm{M^2},\rm{s_0})}{\rho_i (\rm{M^2},\infty)} > 35\%,~~
 \end{align}
and
\begin{align}
 \mbox{CVG} =\frac{\rho_i^{\mbox{DN}} (\rm{M^2},\rm{s_0})}{\rho_i (\rm{M^2},\rm{s_0})} < 5\%,
 \end{align}
 where $\rho_i^{\mbox{DN}} (\rm{M^2},\rm{s_0})$ is  the highest dimensional term in the OPE of $\rho_i (\rm{M^2},\rm{s_0})$. 
On the OPE side of the analysis, the highest-dimensional contribution 
arises from the dimension-7 term 
$\langle g_s^2 G^2 \rangle \langle \bar q q \rangle$. 
Accordingly, the convergence (CVG) analysis has been performed by 
including the D7 contribution, and the results presented in 
Table~\ref{parameter} reflect this choice. In addition, the QCD 
representation incorporates various condensate structures, such as 
$\langle g_s^2 G^2 \rangle f_{3\gamma}$ (D6), 
$\langle g_s^2 G^2 \rangle \langle \bar q q \rangle \chi$ (D5), 
$\langle \bar q q \rangle$ (D3), $f_{3\gamma}$ (D2), and 
$\langle \bar q q \rangle \chi$ (D1). Based on these considerations, the 
working intervals for the auxiliary parameters $\rm{M^2}$ and $\rm{s_0}$ 
have been determined, and the corresponding values of PC and CVG for each 
state are also summarized in Table~\ref{parameter}. 
Figure~\ref{Msqfig1} shows that the PC remains 
sufficiently larger than the continuum contribution within the selected 
Borel window, ensuring the dominance of the ground-state signal. 
At the same time, the relative weights of the condensate terms confirm 
the satisfactory convergence of the OPE expansion. For completeness, 
Fig.~\ref{Msqfig1} illustrates the dependence of the magnetic dipole 
moment of the $D^{(\ast)} \Lambda_c^{(\ast)}$ pentaquarks on the Borel 
mass parameter $\rm{M^2}$ and the threshold parameter $\rm{s_0}$. 
It can be seen from this figure that the dependence on $\rm{M^2}$ is 
rather mild, while the variation with respect to $\rm{s_0}$ is more 
pronounced but still remains within the error margins of the method.

\begin{widetext}

\begin{table}[htb!]
	\addtolength{\tabcolsep}{10pt}
	\caption{Working regions of $\rm{s_0}$ and $\rm{M^2}$, along with PC and CVG, for the magnetic dipole moments of the $D^{(\ast)} \Lambda_c^{(\ast)}$ pentaquarks. }
	\label{parameter}
	\begin{ruledtabular}
\begin{tabular}{l|cccccc}
             \\
States & $\rm{s_0}$ [GeV$^2$] & $\rm{M^2}$ [GeV$^2$]&    PC [$\%$]  &   CVG [$\%$] &  $\mu~[\mu_N]$ \\
 \\
                                        \hline\hline
$ D \Lambda_c$ & $22.0-24.0$ & $2.2-2.8$ & $67.07-41.46$ &  $0.05-0.07$ & $ -1.27 \pm 0.30$
                        \\
$ D \Lambda_c^\ast$& $24.0-26.0$ & $2.2-2.8$ & $67.57-41.20$ &  $0.20-0.22$ & $ -2.78 \pm 0.61$  
                       \\
$ D^\ast \Lambda_c$ & $25.0-27.0$ & $2.4-3.0$ & $62.23-37.71$ &  $0.14-0.18$  & $-3.80 \pm 0.81$ 
                       \\
 \end{tabular}
\end{ruledtabular}
\end{table}

\end{widetext}

Following the numerical evaluations, the resulting predictions for the 
magnetic dipole moments of the $D^{(\ast)} \Lambda_c^{(\ast)}$ pentaquark 
states are given in Table \ref{parameter}.  
The uncertainties for the magnetic moments are obtained by quadratically adding the individual errors originating from the variation of the auxiliary parameters $\mathrm{M}^2$ and $\mathrm{s}_0$ within their working regions, and the uncertainties of the input parameters listed in Table~\ref{inputparameter} and those entering the photon DAs. The dominant source of error is found to be the variation of the continuum threshold $\rm{s_0}$, followed by the uncertainties in the pole residues and hadronic masses.
 
The magnetic dipole moments of the $D^{(\ast)} \Lambda_c^{(\ast)}$ pentaquarks are primarily determined by their internal configurations and the distribution of light and heavy quarks. Among the considered states, the $D \Lambda_c^\ast$ pentaquark exhibits the largest moment due to the specific arrangement of quarks within the spin-3/2 $\Lambda_c^\ast$ baryon, which allows a more constructive combination of individual quark contributions. In contrast, the $D \Lambda_c$ state shows partial cancellations between the light and charm quark contributions, reflecting its spin-1/2 baryon configuration. The $D^\ast \Lambda_c$ state, with a spin-1 meson coupled to a spin-1/2 baryon, exhibits an intermediate behavior, consistent with the interplay of its molecular diquark structure.

A detailed examination of Table~\ref{parameter2} shows that light quarks ($u$ and $d$) provide the dominant contributions to the magnetic moments, while charm quarks contribute significantly but are generally smaller. Interestingly, for the $D \Lambda_c^\ast$ state, the contributions of the light quarks partially cancel, making the charm quark contribution more relevant. This emphasizes that the total magnetic dipole moment is sensitive not only to the spin of the constituent baryon but also to the interplay between quark flavors and the molecular configuration of the pentaquark. 
A nuanced analysis of these contributions reveals distinct mechanisms governing the magnetic moments of each state. For the $J^P = \frac{1}{2}^-$ and $\frac{3}{2}^-$ states ($D\Lambda_c$ and $D^*\Lambda_c$), the charm quark plays a critical role in determining the final value of the magnetic moment by partially canceling the large, opposing contributions from the light quarks. In stark contrast, the mechanism for the $J^P = \frac{3}{2}^-$ $D\Lambda_c^*$ state is qualitatively different. Here, the net light quark contribution and the charm quark contribution are aligned in sign, leading to a constructive addition. Consequently, the charm quark, despite its smaller absolute value, contributes cooperatively to the large negative moment. This stark contrast in the underlying mechanisms—cancellation versus constructive addition—powerfully demonstrates that the electromagnetic properties of these molecular pentaquarks are exquisitely sensitive to the specific spin and isospin configurations of their constituent hadrons. 
It should be noted that the individual quark contributions ($\mu_u, \,\mu_d, \,  \mu_c$) presented in Table~\ref{parameter2} are extracted by isolating the terms proportional to the electric charges $e_u$, $e_d$, and $e_c$ in the sum rules, respectively. They represent the effective contributions within the context of the employed molecular current and should not be interpreted as the magnetic moments of bare quarks.

 
    \begin{table}[htb!]
	\addtolength{\tabcolsep}{10pt}
	\caption{Light and heavy quark contributions to the total magnetic moments of the $D^{(\ast)} \Lambda_c^{(\ast)}$ pentaquark states (in units of nuclear magneton $\mu_N$), where $\mu_{total}= \mu_u +\mu_d +\mu_c$. The results were obtained by utilizing the central values of all input parameters.}
	\label{parameter2}
		\begin{center}
\begin{tabular}{l|ccccc}
	   \hline\hline
	   \\
   States& $\mu_{u}$& $\mu_{d}$&  $\mu_{c}$& $\mu_{total}$\\
  \\
\hline\hline
$ D \Lambda_c$& $-9.10$& $ ~~4.55 $  &  $~~3.28 $& $-1.27$\\
$D^\ast \Lambda_c$& $-13.48 $ & $~~6.74$ &  $~~3.96 $& $-2.78$\\
$D \Lambda_c^\ast $& $~~0.39 $ & $-3.76$ & $-0.43 $& $-3.80$\\
	   \hline\hline
\end{tabular}
\end{center}
\end{table}


Experimentally, direct spin-precession measurements are impractical due to the short lifetimes of $D^{(\ast)} \Lambda_c^{(\ast)}$ pentaquarks. However, their electromagnetic properties can be indirectly probed via radiative transitions in photoproduction or electroproduction processes, exploiting the soft photon emission mechanism~\cite{Zakharov:1968fb, Pascalutsa:2004je}. In these processes, the magnetic dipole moment enters as the leading-order contribution in the expansion of the transition amplitude with respect to the photon energy, while higher multipole contributions are suppressed. By comparing the differential and total cross sections of such radiative processes with theoretical predictions, it should be possible to extract information on the magnetic moments.  The predicted magnetic dipole moments are sizeable and are of the same order of magnitude as the nucleon magnetic moments ($\mu_p \approx 2.79 \mu_N$, $\mu_n \approx -1.91 \mu_N$), which suggests that they could lead to observable effects in radiative processes.

To the best of our knowledge, there are no previous calculations of the magnetic moments for $D^{(\ast)} \Lambda_c^{(\ast)}$ pentaquark configurations. Our predictions thus serve as a crucial first benchmark for future theoretical studies. A particularly revealing comparison can be made with existing predictions for other pentaquark systems obtained using the same QCD light-cone sum rule approach, allowing us to disentangle the effects of quark content, spin-parity, and internal structure.

We first compare our results with doubly-charmed compact pentaquark configurations studied in~\cite{Ozdem:2022vip}. Although these states have a similar five-quark content as our molecular configurations, they are organized into a compact diquark-diquark-antiquark structure rather than a loosely-bound meson-baryon molecule. For the $J^P = 1/2^-$ sector, the compact doubly-charmed pentaquark with quark content $ccud\bar{d}$ yields $\mu_{P_{cc}^{1/2}} = 0.93^{+0.32}_{-0.29}~\mu_N$, which differs dramatically from our $D\Lambda_c$ molecular result ($-1.27 \pm 0.30~\mu_N$) in both sign and magnitude. In the $J^P = 3/2^-$ sector, the compact doubly-charmed configuration with quark content $ccud\bar{d}$ yields $\mu_{P_{cc}^{3/2}} = 2.01^{+0.86}_{-0.75}~\mu_N$. This positive value stands in stark contrast to our $D^*\Lambda_c$ ($-2.78 \pm 0.61~\mu_N$) and $D\Lambda_c^*$ ($-3.80 \pm 0.81~\mu_N$) results, both of which are negative and substantially larger in magnitude. This striking difference demonstrates that even with identical five-quark content ($ccud\bar{d}$), the magnetic moment is exquisitely sensitive to whether the quarks are organized into a compact diquark structure or a molecular meson-baryon configuration.

Beyond this direct comparison with doubly-charmed compact states, it is also instructive to examine how our results relate to hidden-charm pentaquarks. For the $\bar{D}\Lambda_c$ and $\bar{D}^*\Lambda_c$ molecular states, previous calculations yield $\mu_{\bar{D}\Lambda_c} = 0.44^{+0.17}_{-0.14}~\mu_N$ and $\mu_{\bar{D}^*\Lambda_c} = 2.24^{+0.77}_{-0.64}~\mu_N$, respectively~\cite{Ozdem:2023htj}. These hidden-charm molecular states exhibit positive magnetic moments, in contrast to the negative moments we obtain for all three doubly-charmed molecular states. This sign reversal underscores the significant impact of replacing the light antiquark with a second charm quark. 
Compact diquark-diquark-antiquark interpretations of hidden-charm pentaquarks have also been investigated, providing additional points of comparison. The $P_c(4312)$ state, with $J^P = \frac{1}{2}^-$, yields $\mu_{P_c(4312)} = 0.40 \pm 0.15~\mu_N$ in the compact picture~\cite{Ozdem:2021btf}. For the $P_c(4337)$ pentaquark, studied under the same compact framework, the magnetic moment is found to be $\mu_{P_c(4337)} = 1.76 \pm 0.44~\mu_N$ for $J^P = \frac{1}{2}^-$ and $\mu_{P_c(4337)} = -1.38 \pm 0.35~\mu_N$ for $J^P = \frac{3}{2}^-$~\cite{Ozdem:2025fks}. Meanwhile, the $P_c(4457)$ state in the compact diquark-diquark-antiquark picture gives $\mu_{P_c(4457)} = -1.96 \pm 0.50~\mu_N$ for $J^P = \frac{3}{2}^-$~\cite{Ozdem:2024suc}.  
When compared with these hidden-charm compact configurations, our $D\Lambda_c$ result ($-1.27 \pm 0.30~\mu_N$) differs markedly from the positive $P_c(4312)$ compact value, while our $D^*\Lambda_c$ result ($-2.78 \pm 0.61~\mu_N$) shows better agreement in sign with the $J^P = \frac{3}{2}^-$ assignment of $P_c(4337)$ ($-1.38 \pm 0.35~\mu_N$). Our $D\Lambda_c^*$ result ($-3.80 \pm 0.81~\mu_N$) aligns in sign with the $P_c(4457)$ compact prediction ($-1.96 \pm 0.50~\mu_N$), though with a significantly larger magnitude. These systematic comparisons are summarized in Table~\ref{tab:complete-comparison}.

\begin{table}[htbp]
\centering
\caption{Comparison of our molecular doubly-charmed results with compact doubly-charmed and hidden-charm configurations.}
\begin{tabular}{lccc}
\hline
\hline
\\
States & $\mu$ [$\mu_N$] & Compact counterpart & $\mu$ [$\mu_N$] \\
\\
\hline \hline
\multirow{2}{*}{$D\Lambda_c$ } & \multirow{2}{*}{$-1.27 \pm 0.30$} & $P_{cc}^{1/2}$~\cite{Ozdem:2022vip} & $0.93^{+0.32}_{-0.29}$ \\
 & & $P_c(4312)$ ~\cite{Ozdem:2021btf} & $~~0.40 \pm 0.15$ \\
\hline
\multirow{2}{*}{$D^*\Lambda_c$} & \multirow{2}{*}{$-2.78 \pm 0.61$} & $P_{cc}^{3/2}$~\cite{Ozdem:2022vip} & $2.01^{+0.86}_{-0.75}$ \\
 & & $P_c(4337)$~\cite{Ozdem:2025fks} & $-1.38 \pm 0.35$ \\
\hline
\multirow{2}{*}{$D\Lambda_c^*$ } & \multirow{2}{*}{$-3.80 \pm 0.81$} & $P_{cc}^{3/2}$~\cite{Ozdem:2022vip} & $2.01^{+0.86}_{-0.75}$ \\
 & & $P_c(4457)$~\cite{Ozdem:2024suc} & $-1.96 \pm 0.50$ \\
\hline
\hline
\end{tabular}
\label{tab:complete-comparison}
\end{table}

The comprehensive comparison presented above reveals several important patterns. First, comparing our molecular doubly-charmed results with compact doubly-charmed configurations of similar five-quark content demonstrates that the magnetic moment is not merely determined by which quarks are present, but critically depends on how they are organized---whether as a compact diquark structure or a loosely-bound meson-baryon molecule. This sign reversal and changes in magnitude underscore the dramatic impact of charm quark content on the electromagnetic structure. Second, the comparison with hidden-charm systems, both molecular and compact, reveals the profound impact of charm quark count. Replacing the light antiquark in hidden-charm systems with a second charm quark in doubly-charmed states systematically changes the magnetic response and, in most cases, reverses its sign. This demonstrates that electromagnetic observables are exquisitely sensitive to the specific charm quark distribution within the five-quark system.

The relatively large predicted magnetic moments in our doubly-charmed systems, combined with these distinctive sign patterns across different configurations, provide a clear experimental signature for distinguishing between different charm quark configurations and internal structures. These findings will be essential for the eventual experimental identification of these states through their electromagnetic properties. The marked differences between molecular and compact organizations, as well as between hidden-charm and doubly-charmed systems, revealed through the consistent application of QCD light-cone sum rules, highlight the rich diversity of electromagnetic behavior in exotic hadrons and offer a powerful probe of their internal quark-gluon dynamics.

Having extended our analysis beyond magnetic dipole moments to higher electromagnetic multipoles, our framework predicts the electric quadrupole moment $\mathcal{Q}$ and magnetic octupole moment $\mathcal{O}$ for the $D^\ast \Lambda_c$ and $D \Lambda_c^\ast$ molecular states, providing crucial insights into their spatial deformation characteristics. The results are as follows:
\begin{align}
& \mathcal Q_{D^\ast \Lambda_c} = (2.73 \pm 0.45) \times 10^{-2}  \mathrm{fm}^2,~~~~~~~\nonumber \\
&  \mathcal O_{D^\ast \Lambda_c} = (0.55 \pm 0.08) \times 10^{-3}  \mathrm{fm}^3,
 \end{align}
 \begin{align}
& \mathcal Q_{D \Lambda_c^\ast} = (-2.07 \pm 0.60) \times 10^{-2}  \mathrm{fm}^2, ~~~~~\nonumber\\ 
& \mathcal O_{D \Lambda_c^\ast} = (-0.48 \pm 0.12) \times 10^{-3}  \mathrm{fm}^3. 
\end{align}

These values reveal a clear physical pattern: the $D^\ast \Lambda_c$ state exhibits a small but positive quadrupole moment, suggesting a prolate (cigar-like) deformation of its charge distribution. In contrast, the $D \Lambda_c^\ast$ state has a negative quadrupole moment, consistent with an oblate (pancake-like) shape. The small magnitudes of both $\mathcal{Q}$ and $\mathcal{O}$ reflect the molecular nature of these states, where the multipole moments arise from the loosely bound system as a whole, rather than from a single compact particle.

To gain deeper insight into the underlying mechanisms responsible for these deformation patterns, we decompose the multipole moments into individual quark contributions. As shown in Table~\ref{parameter3}, this decomposition reveals distinct dynamics governing the electromagnetic structure of each state.  
For the $D^\ast \Lambda_c$ state, the positive total quadrupole moment ($\mathcal{Q} = 2.73$) arises from a constructive interplay where the dominant positive contribution from the $u$-quark ($\mathcal{Q}_u = 3.62$) is partially canceled by the negative $d$-quark contribution ($\mathcal{Q}_d = -1.81$), with the charm quark providing a smaller positive addition ($\mathcal{Q}_c = 0.92$). This pattern suggests a prolate deformation primarily driven by light quark dynamics, with the charm quark reinforcing the overall shape. The magnetic octupole moment follows the same trend, showing aligned signs across all quark contributions. 
In stark contrast, the $D \Lambda_c^\ast$ state exhibits a qualitatively different mechanism. Here, the charm quark dominates with a large negative contribution ($\mathcal{Q}_c = -2.92$) that determines the oblate character of the system ($\mathcal{Q} = -2.07$). The light quarks provide small positive contributions ($\mathcal{Q}_u = 0.25$, $\mathcal{Q}_d = 0.60$) that partially counteract the charm quark effect but are insufficient to change the overall negative sign. This emphasizes that the deformation properties are sensitive not only to the molecular configuration but also to which constituent baryon carries the excited charm degree of freedom. The magnetic octupole moments mirror these patterns, confirming that the higher multipole structure of these states is governed by the same underlying quark-level dynamics. The consistent alignment between quadrupole and octupole moments for each state provides crucial fingerprints for understanding their internal structure and deformation characteristics. These predicted patterns offer clear benchmarks for future spectroscopic studies aiming to confirm the molecular nature of these exotic states.

The combination of quantitative multipole moments in Tables~\ref{parameter2}--\ref{parameter3} with the qualitative 3D visualizations in Figs.~\ref{quadrupole3D}--\ref{octupole3D} provides a comprehensive picture of the electromagnetic structure of these pentaquark states. The clear correspondence between the calculated moment values and the observed deformation patterns serves as a robust consistency check of our results. 
These visualizations particularly enhance our understanding of the deformation mechanisms, revealing how the quadrupole moments reflect the overall shape anisotropy while the octupole moments capture more subtle asymmetries in the charge and current distributions. These distinctive multipole signatures could serve as crucial benchmarks in future experimental observations of radiative transitions.

\begin{widetext}

    \begin{table}[htb!]
	\addtolength{\tabcolsep}{10pt}
	\caption{Individual light and heavy quark contributions to the electric quadrupole ($\mathcal{Q}$) and magnetic octupole ($\mathcal{O}$) moments of the $D^\ast \Lambda_c$ and $D \Lambda_c^\ast$ pentaquark states. All quadrupole moments are in units of $10^{-2}~\mathrm{fm}^2$ and octupole moments in units of $10^{-3}~\mathrm{fm}^3$.}
	\label{parameter3}
\begin{center}
\begin{tabular}{l|cccc|cccc}
\hline\hline
\multirow{2}{*}{States} & \multicolumn{4}{c|}{Electric Quadrupole Moment, $\mathcal{Q}$} & \multicolumn{4}{c}{Magnetic Octupole Moment, $\mathcal{O}$} \\
		\cline{2-9}
& $\mathcal{Q}_u$ & $\mathcal{Q}_d$ & $\mathcal{Q}_c$ & $\mathcal{Q}_{\text{total}}$ & $\mathcal{O}_u$ & $\mathcal{O}_d$ & $\mathcal{O}_c$ & $\mathcal{O}_{\text{total}}$ \\
\hline\hline
$D^\ast \Lambda_c$    & $ 3.62$ & $-1.81$ & $ ~~0.92$ & $ ~~2.73$ & $ 0.68$ & $-0.34$ & $~~ 0.21$ & $~~ 0.55$ \\
$D \Lambda_c^\ast$    & $ 0.25$ & $~~ 0.60$ & $-2.92$ & $-2.07$ & $ 0.04$ & $ ~~0.06$ & $-0.58$ & $-0.48$ \\
\hline\hline
\end{tabular}
\end{center}
	\end{table}
	
\end{widetext}

To conclude, the present analysis is performed assuming that the investigated doubly-charmed pentaquarks are predominantly molecular states composed of color-singlet meson–baryon constituents. It is therefore instructive to comment on how the predicted magnetic dipole moments may depend on the assumed internal structure and on the strength of the molecular binding. If a compact multiquark configuration were assumed instead, one would expect a qualitatively different electromagnetic response. In compact pentaquark scenarios, quarks are correlated at shorter distances with nontrivial color structures, such as diquark–triquark or diquark–diquark–antiquark configurations. In such cases, heavy-quark spin correlations are typically enhanced, and the contribution of the charm quarks to the magnetic dipole moment is expected to be more pronounced (see, for example, Refs.~\cite{Ozdem:2024rqx, Ozdem:2024rch, Ozdem:2025ncd}). As a consequence, both the magnitude and even the sign of the magnetic dipole moments could differ substantially from the molecular predictions obtained in this work. In contrast, the molecular configuration adopted here favors a long-range structure in which the electromagnetic response is largely governed by the light-quark degrees of freedom, leading to the observed negative magnetic moments and their characteristic hierarchy. Regarding the sensitivity to the molecular binding, it should be emphasized that within the QCD light-cone sum rule framework binding effects enter indirectly through the hadronic masses and pole residues. Variations in the binding energy primarily modify these input parameters, which are already varied within their uncertainties in the numerical analysis. We find that the resulting magnetic dipole moments exhibit only a mild dependence on such variations, indicating that the dominant sensitivity originates from the underlying spin–flavor structure of the molecular configuration rather than from the precise value of the binding energy. The possible presence of non-molecular components can be interpreted in terms of an effective molecular probability. A reduced molecular probability would imply an increasing admixture of compact components, for which a qualitatively different electromagnetic behavior is expected. Consequently, future experimental or theoretical deviations from our predictions could signal the relevance of compact configurations beyond a purely molecular picture. It is worth noting that electromagnetic observables such as magnetic dipole moments are particularly sensitive to the assumed internal configuration of multiquark states. Although physical observables are in principle  unique, quantities closely tied to internal dynamics—such as electromagnetic multipole moments—can exhibit significant variations depending on the underlying structural picture, as demonstrated in several studies of multiquark systems~\cite{Wang:2016dzu, Ozdem:2024rch, Gao:2021hmv, Li:2024wxr, Li:2024jlq, Ozdem:2024rqx, Ozdem:2024txt, Mutuk:2024ach, Ozdem:2025jda, Ozdem:2024dbq, Ozdem:2024lpk, Azizi:2023gzv}. In this sense, the magnetic dipole moments predicted in this work provide robust benchmarks for distinguishing between different internal configurations of doubly-charmed pentaquarks.  

  \begin{widetext}

\begin{figure}[htb!]
\centering
\includegraphics[width=1.0\textwidth]{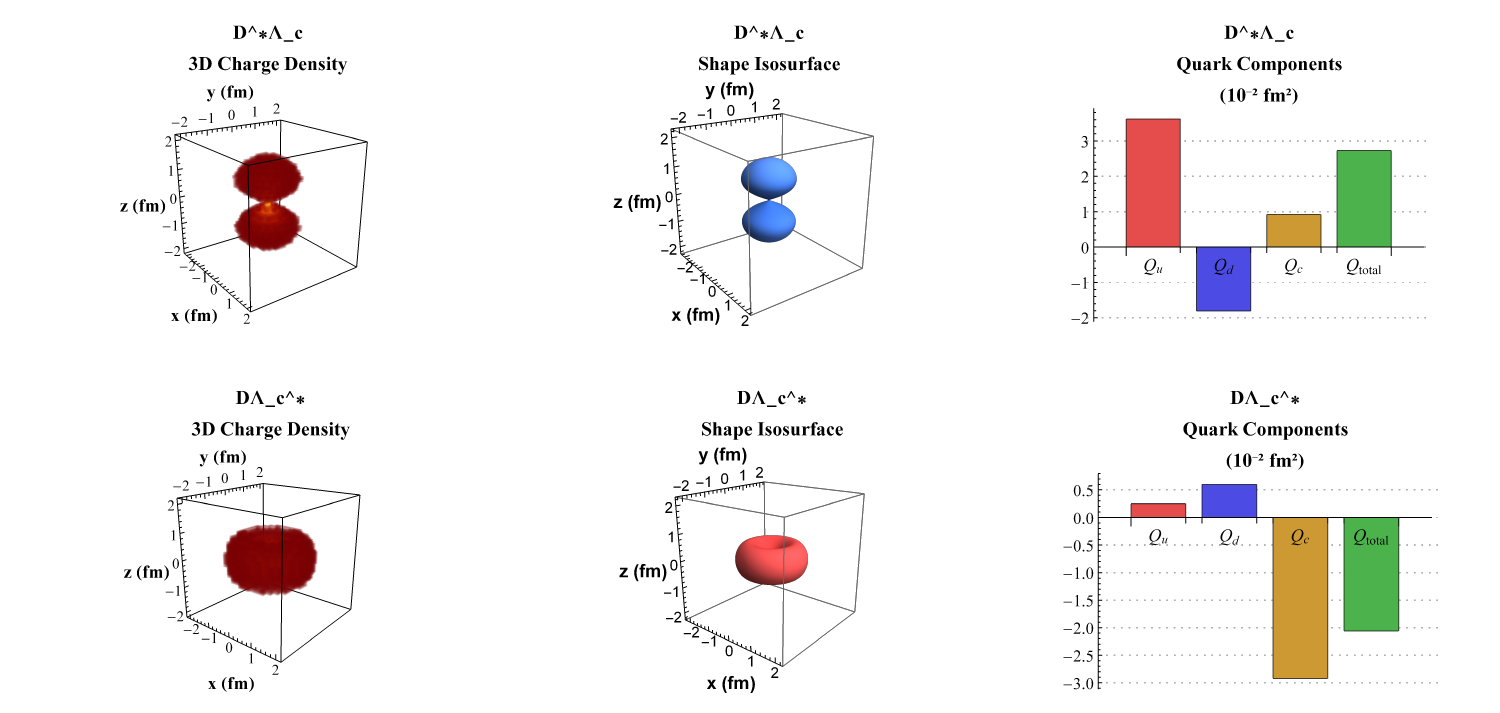}
 \caption{Analysis of the electric quadrupole moment for the $D^\ast \Lambda_c$ and $D \Lambda_c^\ast$ pentaquark states. Left: three-dimensional charge density distribution; Middle: isosurface visualization showing deformation patterns; Right: individual quark contributions to the quadrupole moment. The plots reveal distinct prolate and oblate deformations characteristic of each pentaquark configuration, with color coding indicating density intensity. All spatial axes are in femtometers (fm).}
 \label{quadrupole3D}
  \end{figure}


\begin{figure}[htb!]
\centering
\includegraphics[width=1.0\textwidth]{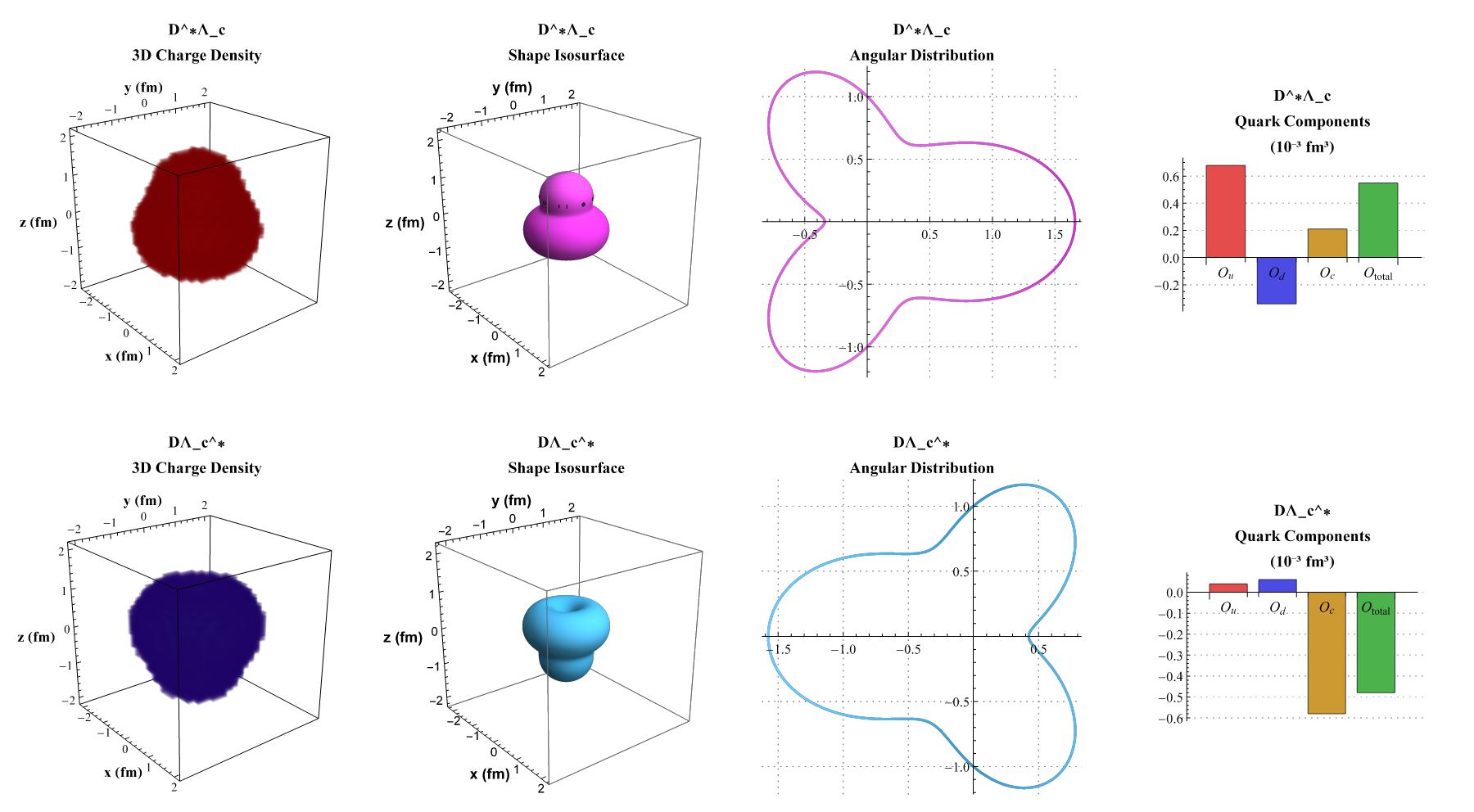}
 \caption{Analysis of the magnetic octupole moment for the $D^\ast \Lambda_c$ and $D \Lambda_c^\ast$ pentaquark states. Left: three-dimensional charge density distribution; Middle: isosurface visualization showing higher-order deformations; Right: angular distribution profile; Far right: individual quark contributions to the octupole moment. The plots reveal peanut-shaped and butterfly-shaped deformations characteristic of positive and negative octupole moments, with color coding indicating density intensity. All spatial axes are in femtometers (fm).}
 \label{octupole3D}
  \end{figure}

 \end{widetext}

\section{Summary of results}\label{summary}

In this work, we have investigated the magnetic dipole moments of the $D^{(\ast)} \Lambda_c^{(\ast)}$ pentaquarks assuming molecular configurations $D \Lambda_c$, $D^\ast \Lambda_c$, and $D \Lambda_c^\ast$ with quantum numbers $J^P = \frac{1}{2}^-$, $\frac{3}{2}^-$, and $\frac{3}{2}^-$, respectively, using QCD light-cone sum rules. The analysis includes contributions from both light and charm quarks, as well as nonperturbative effects from quark and gluon condensates and photon distribution amplitudes.

Our results show that the magnetic dipole moments of the $D^{(\ast)} \Lambda_c^{(\ast)}$ pentaquarks are mainly determined by their internal diquark–quark structure and the arrangement of light and heavy quarks. The $D \Lambda_c^\ast$ state has the largest moment, reflecting a configuration in which the quark contributions add constructively, whereas in the $D \Lambda_c$ state, partial cancellations among quark contributions reduce the total moment. The $D^\ast \Lambda_c$ state exhibits intermediate behavior, consistent with its molecular diquark arrangement. In general, light quarks provide the dominant contribution, while charm quarks play a smaller but non-negligible role. This is particularly evident in the $D \Lambda_c^\ast$ state, where cancellations among light quark contributions enhance the relative importance of the charm quark.

Beyond the dipole moments, our investigation of higher electromagnetic multipoles—electric quadrupole and magnetic octupole moments—for the $D^*\Lambda_c$ and $D\Lambda_c^*$ states reveals distinctive deformation patterns in their charge distributions. The electric quadrupole and magnetic octupole moments provide crucial insights into the spatial structure of the pentaquarks. The positive quadrupole moment of the $D^\ast \Lambda_c$ state indicates a prolate (cigar-shaped) deformation, while the negative value for $D \Lambda_c^\ast$ suggests an oblate (pancake-like) shape. The decomposition into quark contributions shows that these deformation patterns arise from fundamentally different mechanisms: light-quark dominance in $D^\ast \Lambda_c$ versus charm-quark dominance in $D \Lambda_c^\ast$. The small magnitudes of all multipole moments further support the molecular nature of these states.

These findings suggest that electromagnetic observables, such as the magnetic dipole moments and higher multipoles, provide a sensitive probe of the internal quark-gluon dynamics and the molecular nature of doubly-charmed pentaquarks. While direct experimental measurement is challenging due to the short lifetimes of these states, radiative transitions in photoproduction or electroproduction processes offer a promising indirect approach. The predicted multipole patterns serve as specific fingerprints that could help identify these states in future spectroscopic studies. Overall, our study highlights the potential of electromagnetic moments to distinguish between molecular and more compact multiquark configurations in future investigations.

\bibliographystyle{elsarticle-num}
\bibliography{DLambdacMDM.bib}

@article{Aaij:2015tga,
    author = "Aaij, Roel and others",
    collaboration = "LHCb",
    title = "{Observation of $J/\psi p$ Resonances Consistent with Pentaquark States in $\Lambda_b^0 \to J/\psi K^- p$ Decays}",
    eprint = "1507.03414",
    archivePrefix = "arXiv",
    primaryClass = "hep-ex",
    reportNumber = "CERN-PH-EP-2015-153, LHCB-PAPER-2015-029",
    doi = "10.1103/PhysRevLett.115.072001",
    journal = "Phys. Rev. Lett.",
    volume = "115",
    pages = "072001",
    year = "2015"
}

@article{Aaij:2019vzc,
    author = "Aaij, Roel and others",
    collaboration = "LHCb",
    title = "{Observation of a narrow pentaquark state, $P_c(4312)^+$, and of two-peak structure of the $P_c(4450)^+$}",
    eprint = "1904.03947",
    archivePrefix = "arXiv",
    primaryClass = "hep-ex",
    reportNumber = "LHCb-PAPER-2019-014 CERN-EP-2019-058",
    doi = "10.1103/PhysRevLett.122.222001",
    journal = "Phys. Rev. Lett.",
    volume = "122",
    number = "22",
    pages = "222001",
    year = "2019"
}

@article{LHCb:2021chn,
    author = "Aaij, Roel and others",
    collaboration = "LHCb",
    title = "{Evidence for a new structure in the $J/\psi p$ and $J/\psi \bar{p}$ systems in $B_s^0 \to J/\psi p \bar{p}$ decays}",
    eprint = "2108.04720",
    archivePrefix = "arXiv",
    primaryClass = "hep-ex",
    reportNumber = "LHCb-PAPER-2021-018, CERN-EP-2021-150",
    doi = "10.1103/PhysRevLett.128.062001",
    journal = "Phys. Rev. Lett.",
    volume = "128",
    number = "6",
    pages = "062001",
    year = "2022"
}

@article{Collaboration:2022boa,
    author = "Aaij, R. and others",
    collaboration = "LHCb",
    title = "{Observation of a J/{\ensuremath{\psi}}{\ensuremath{\Lambda}} Resonance Consistent with a Strange Pentaquark Candidate in B-{\textrightarrow}J/{\ensuremath{\psi}}{\ensuremath{\Lambda}}p{\textasciimacron} Decays}",
    eprint = "2210.10346",
    archivePrefix = "arXiv",
    primaryClass = "hep-ex",
    reportNumber = "CERN-EP-2022-198, LHCb-PAPER-2022-031",
    doi = "10.1103/PhysRevLett.131.031901",
    journal = "Phys. Rev. Lett.",
    volume = "131",
    number = "3",
    pages = "031901",
    year = "2023"
}

@article{LHCb:2021vvq,
    author = "Aaij, Roel and others",
    collaboration = "LHCb",
    title = "{Observation of an exotic narrow doubly charmed tetraquark}",
    eprint = "2109.01038",
    archivePrefix = "arXiv",
    primaryClass = "hep-ex",
    reportNumber = "CERN-EP-2021-165, LHCb-PAPER-2021-031",
    doi = "10.1038/s41567-022-01614-y",
    journal = "Nature Phys.",
    volume = "18",
    number = "7",
    pages = "751--754",
    year = "2022"
}

@article{LHCb:2021auc,
    author = "Aaij, Roel and others",
    collaboration = "LHCb",
    title = "{Study of the doubly charmed tetraquark $T_{cc}^{+}$}",
    eprint = "2109.01056",
    archivePrefix = "arXiv",
    primaryClass = "hep-ex",
    reportNumber = "CERN-EP-2021-169, LHCb-PAPER-2021-032",
    doi = "10.1038/s41467-022-30206-w",
    journal = "Nature Commun.",
    volume = "13",
    number = "1",
    pages = "3351",
    year = "2022"
}

@article{Esposito:2014rxa,
    author = "Esposito, Angelo and Guerrieri, Andrea L. and Piccinini, Fulvio and Pilloni, Alessandro and Polosa, Antonio D.",
    title = "{Four-Quark Hadrons: an Updated Review}",
    eprint = "1411.5997",
    archivePrefix = "arXiv",
    primaryClass = "hep-ph",
    doi = "10.1142/S0217751X15300021",
    journal = "Int. J. Mod. Phys. A",
    volume = "30",
    pages = "1530002",
    year = "2015"
}

@article{Esposito:2016noz,
    author = "Esposito, A. and Pilloni, A. and Polosa, A. D.",
    title = "{Multiquark Resonances}",
    eprint = "1611.07920",
    archivePrefix = "arXiv",
    primaryClass = "hep-ph",
    reportNumber = "JLAB-THY-16-2301",
    doi = "10.1016/j.physrep.2016.11.002",
    journal = "Phys. Rept.",
    volume = "668",
    pages = "1--97",
    year = "2017"
}

@article{Olsen:2017bmm,
    author = "Olsen, Stephen Lars and Skwarnicki, Tomasz and Zieminska, Daria",
    title = "{Nonstandard heavy mesons and baryons: Experimental evidence}",
    eprint = "1708.04012",
    archivePrefix = "arXiv",
    primaryClass = "hep-ph",
    doi = "10.1103/RevModPhys.90.015003",
    journal = "Rev. Mod. Phys.",
    volume = "90",
    number = "1",
    pages = "015003",
    year = "2018"
}

@article{Lebed:2016hpi,
    author = "Lebed, Richard F. and Mitchell, Ryan E. and Swanson, Eric S.",
    title = "{Heavy-Quark QCD Exotica}",
    eprint = "1610.04528",
    archivePrefix = "arXiv",
    primaryClass = "hep-ph",
    doi = "10.1016/j.ppnp.2016.11.003",
    journal = "Prog. Part. Nucl. Phys.",
    volume = "93",
    pages = "143--194",
    year = "2017"
}

@article{Nielsen:2009uh,
    author = "Nielsen, Marina and Navarra, Fernando S. and Lee, Su Houng",
    title = "{New Charmonium States in QCD Sum Rules: A Concise Review}",
    eprint = "0911.1958",
    archivePrefix = "arXiv",
    primaryClass = "hep-ph",
    doi = "10.1016/j.physrep.2010.07.005",
    journal = "Phys. Rept.",
    volume = "497",
    pages = "41--83",
    year = "2010"
}

@article{Brambilla:2019esw,
    author = "Brambilla, Nora and Eidelman, Simon and Hanhart, Christoph and Nefediev, Alexey and Shen, Cheng-Ping and Thomas, Christopher E. and Vairo, Antonio and Yuan, Chang-Zheng",
    title = "{The $XYZ$ states: experimental and theoretical status and perspectives}",
    eprint = "1907.07583",
    archivePrefix = "arXiv",
    primaryClass = "hep-ex",
    reportNumber = "TUM-EFT 125/19",
    doi = "10.1016/j.physrep.2020.05.001",
    journal = "Phys. Rept.",
    volume = "873",
    pages = "1--154",
    year = "2020"
}

@article{Agaev:2020zad,
    author = "Agaev, Shahin and Azizi, Kazem and Sundu, Hayriye",
    title = "{Four-quark exotic mesons}",
    eprint = "2004.12079",
    archivePrefix = "arXiv",
    primaryClass = "hep-ph",
    doi = "10.3906/fiz-2003-15",
    journal = "Turk. J. Phys.",
    volume = "44",
    number = "2",
    pages = "95--173",
    year = "2020"
}

@article{Chen:2016qju,
    author = "Chen, Hua-Xing and Chen, Wei and Liu, Xiang and Zhu, Shi-Lin",
    title = "{The hidden-charm pentaquark and tetraquark states}",
    eprint = "1601.02092",
    archivePrefix = "arXiv",
    primaryClass = "hep-ph",
    doi = "10.1016/j.physrep.2016.05.004",
    journal = "Phys. Rept.",
    volume = "639",
    pages = "1--121",
    year = "2016"
}

@article{Ali:2017jda,
    author = {Ali, Ahmed and Lange, Jens S{\"o}ren and Stone, Sheldon},
    title = "{Exotics: Heavy Pentaquarks and Tetraquarks}",
    eprint = "1706.00610",
    archivePrefix = "arXiv",
    primaryClass = "hep-ph",
    reportNumber = "DESY-17-071",
    doi = "10.1016/j.ppnp.2017.08.003",
    journal = "Prog. Part. Nucl. Phys.",
    volume = "97",
    pages = "123--198",
    year = "2017"
}

@article{Guo:2017jvc,
    author = "Guo, Feng-Kun and Hanhart, Christoph and Mei{\ss}ner, Ulf-G. and Wang, Qian and Zhao, Qiang and Zou, Bing-Song",
    title = "{Hadronic molecules}",
    eprint = "1705.00141",
    archivePrefix = "arXiv",
    primaryClass = "hep-ph",
    doi = "10.1103/RevModPhys.90.015004",
    journal = "Rev. Mod. Phys.",
    volume = "90",
    number = "1",
    pages = "015004",
    year = "2018",
    note = "[Erratum: Rev.Mod.Phys. 94, 029901 (2022)]"
}

@article{Liu:2019zoy,
    author = "Liu, Yan-Rui and Chen, Hua-Xing and Chen, Wei and Liu, Xiang and Zhu, Shi-Lin",
    title = "{Pentaquark and Tetraquark states}",
    eprint = "1903.11976",
    archivePrefix = "arXiv",
    primaryClass = "hep-ph",
    doi = "10.1016/j.ppnp.2019.04.003",
    journal = "Prog. Part. Nucl. Phys.",
    volume = "107",
    pages = "237--320",
    year = "2019"
}

@article{Yang:2020atz,
    author = "Yang, Gang and Ping, Jialun and Segovia, Jorge",
    title = "{Tetra- and penta-quark structures in the constituent quark model}",
    eprint = "2009.00238",
    archivePrefix = "arXiv",
    primaryClass = "hep-ph",
    doi = "10.3390/sym12111869",
    journal = "Symmetry",
    volume = "12",
    number = "11",
    pages = "1869",
    year = "2020"
}

@article{Dong:2021juy,
    author = "Dong, Xiang-Kun and Guo, Feng-Kun and Zou, Bing-Song",
    title = "{A survey of heavy-antiheavy hadronic molecules}",
    eprint = "2101.01021",
    archivePrefix = "arXiv",
    primaryClass = "hep-ph",
    doi = "10.13725/j.cnki.pip.2021.02.001",
    journal = "Progr. Phys.",
    volume = "41",
    pages = "65--93",
    year = "2021"
}

@article{Meng:2022ozq,
    author = "Meng, Lu and Wang, Bo and Wang, Guang-Juan and Zhu, Shi-Lin",
    title = "{Chiral perturbation theory for heavy hadrons and chiral effective field theory for heavy hadronic molecules}",
    eprint = "2204.08716",
    archivePrefix = "arXiv",
    primaryClass = "hep-ph",
    doi = "10.1016/j.physrep.2023.04.003",
    journal = "Phys. Rept.",
    volume = "1019",
    pages = "1--149",
    year = "2023"
}

@article{Chen:2022asf,
    author = "Chen, Hua-Xing and Chen, Wei and Liu, Xiang and Liu, Yan-Rui and Zhu, Shi-Lin",
    title = "{An updated review of the new hadron states}",
    eprint = "2204.02649",
    archivePrefix = "arXiv",
    primaryClass = "hep-ph",
    doi = "10.1088/1361-6633/aca3b6",
    journal = "Rept. Prog. Phys.",
    volume = "86",
    number = "2",
    pages = "026201",
    year = "2023"
}

@article{Dong:2021bvy,
    author = "Dong, Xiang-Kun and Guo, Feng-Kun and Zou, Bing-Song",
    title = "{A survey of heavy{\textendash}heavy hadronic molecules}",
    eprint = "2108.02673",
    archivePrefix = "arXiv",
    primaryClass = "hep-ph",
    doi = "10.1088/1572-9494/ac27a2",
    journal = "Commun. Theor. Phys.",
    volume = "73",
    number = "12",
    pages = "125201",
    year = "2021"
}

@article{Duan:2024uuf,
    author = "Duan, Feng-Bo and Wang, Qi-Nan and Yang, Zi-Yan and Chen, Xu-Liang and Chen, Wei",
    title = "{Doubly charmed pentaquark states in QCD sum rules}",
    eprint = "2401.10078",
    archivePrefix = "arXiv",
    primaryClass = "hep-ph",
    doi = "10.1103/PhysRevD.109.094018",
    journal = "Phys. Rev. D",
    volume = "109",
    number = "9",
    pages = "094018",
    year = "2024"
}

@article{Ozdem:2022vip,
    author = {{\"O}zdem, Ulas},
    title = "{Electromagnetic properties of doubly heavy pentaquark states}",
    eprint = "2201.00979",
    archivePrefix = "arXiv",
    primaryClass = "hep-ph",
    doi = "10.1140/epjp/s13360-022-03125-4",
    journal = "Eur. Phys. J. Plus",
    volume = "137",
    pages = "936",
    year = "2022"
}

@article{Yang:2025aer,
    author = "Yang, Zi-Yan and Chen, Wei",
    title = "{Decay and production properties of strange double charm pentaquark}",
    eprint = "2511.04026",
    archivePrefix = "arXiv",
    primaryClass = "hep-ph",
    month = "11",
    year = "2025"
}

@article{Wang:2016dzu,
    author = "Wang, Guang-Juan and Chen, Rui and Ma, Li and Liu, Xiang and Zhu, Shi-Lin",
    title = "{Magnetic moments of the hidden-charm pentaquark states}",
    eprint = "1605.01337",
    archivePrefix = "arXiv",
    primaryClass = "hep-ph",
    doi = "10.1103/PhysRevD.94.094018",
    journal = "Phys. Rev. D",
    volume = "94",
    number = "9",
    pages = "094018",
    year = "2016"
}

@article{Ozdem:2024rch,
    author = {{\"O}zdem, Ula{\c{s}}},
    title = "{Shedding light on the nature of the Pcs(4459) pentaquark state}",
    eprint = "2411.11442",
    archivePrefix = "arXiv",
    primaryClass = "hep-ph",
    doi = "10.1103/PhysRevD.111.074038",
    journal = "Phys. Rev. D",
    volume = "111",
    number = "7",
    pages = "074038",
    year = "2025"
}

@article{Ozdem:2024rqx,
    author = {{\"O}zdem, Ula{\c{s}}},
    title = "{Elucidating the nature of hidden-charm pentaquark states with spin-32 through their electromagnetic form factors}",
    eprint = "2402.03802",
    archivePrefix = "arXiv",
    primaryClass = "hep-ph",
    doi = "10.1016/j.physletb.2024.138551",
    journal = "Phys. Lett. B",
    volume = "851",
    pages = "138551",
    year = "2024"
}

@article{Ozdem:2023htj,
    author = {{\"O}zdem, Ula{\c{s}}},
    title = "{Electromagnetic properties of D{\textasciimacron}({\textasteriskcentered}){\ensuremath{\Xi}}c', D{\textasciimacron}({\textasteriskcentered}){\ensuremath{\Lambda}}c, D{\textasciimacron}s({\textasteriskcentered}){\ensuremath{\Lambda}}c and D{\textasciimacron}s({\textasteriskcentered}){\ensuremath{\Xi}}c pentaquarks}",
    eprint = "2303.10649",
    archivePrefix = "arXiv",
    primaryClass = "hep-ph",
    doi = "10.1016/j.physletb.2023.138267",
    journal = "Phys. Lett. B",
    volume = "846",
    pages = "138267",
    year = "2023"
}

@article{Ozdem:2022kei,
    author = {{\"O}zdem, Ula{\c{s}}},
    title = "{Investigation of magnetic moment of Pcs(4338) and Pcs(4459) pentaquark states}",
    eprint = "2208.07684",
    archivePrefix = "arXiv",
    primaryClass = "hep-ph",
    doi = "10.1016/j.physletb.2022.137635",
    journal = "Phys. Lett. B",
    volume = "836",
    pages = "137635",
    year = "2023"
}

@article{Ozdem:2018qeh,
    author = {{\"O}zdem, U. and Azizi, K.},
    title = "{Electromagnetic multipole moments of the $P_c^+(4380)$ pentaquark in light-cone QCD}",
    eprint = "1803.06831",
    archivePrefix = "arXiv",
    primaryClass = "hep-ph",
    doi = "10.1140/epjc/s10052-018-5873-2",
    journal = "Eur. Phys. J. C",
    volume = "78",
    number = "5",
    pages = "379",
    year = "2018"
}

@article{Ortiz-Pacheco:2018ccl,
    author = "Ortiz-Pacheco, Emmanuel and Bijker, Roelof and Fern{\'a}ndez-Ram{\'\i}rez, C{\'e}sar",
    title = "{Hidden charm pentaquarks: mass spectrum, magnetic moments, and photocouplings}",
    eprint = "1808.10512",
    archivePrefix = "arXiv",
    primaryClass = "nucl-th",
    doi = "10.1088/1361-6471/ab096d",
    journal = "J. Phys. G",
    volume = "46",
    number = "6",
    pages = "065104",
    year = "2019"
}

@article{Xu:2020flp,
    author = "Xu, Yong-Jiang and Liu, Yong-Lu and Huang, Ming-Qiu",
    title = "{The magnetic moment of $P_{c}(4312)$ as a $\bar{D}\Sigma _{c}$ molecular state}",
    eprint = "2008.07937",
    archivePrefix = "arXiv",
    primaryClass = "hep-ph",
    doi = "10.1140/epjc/s10052-021-09211-8",
    journal = "Eur. Phys. J. C",
    volume = "81",
    number = "5",
    pages = "421",
    year = "2021"
}

@article{Ozdem:2021btf,
    author = {{\"O}zdem, Ula{\c{s}}},
    title = "{Electromagnetic properties of the $P_c$ (4312) pentaquark state}",
    doi = "10.1088/1674-1137/abd01c",
    journal = "Chin. Phys. C",
    volume = "45",
    number = "2",
    pages = "023119",
    year = "2021"
}

@article{Ozdem:2021ugy,
    author = {{\"O}zdem, Ula{\c{s}}},
    title = "{Magnetic dipole moments of the hidden-charm pentaquark states: $P_c(4440)$, $P_c(4457)$ and $P_{cs}(4459)$}",
    eprint = "2102.01996",
    archivePrefix = "arXiv",
    primaryClass = "hep-ph",
    doi = "10.1140/epjc/s10052-021-09070-3",
    journal = "Eur. Phys. J. C",
    volume = "81",
    number = "4",
    pages = "277",
    year = "2021"
}

@article{Li:2021ryu,
    author = "Li, Ming-Wei and Liu, Zhan-Wei and Sun, Zhi-Feng and Chen, Rui",
    title = "{Magnetic moments and transition magnetic moments of Pc and Pcs states}",
    eprint = "2106.15053",
    archivePrefix = "arXiv",
    primaryClass = "hep-ph",
    doi = "10.1103/PhysRevD.104.054016",
    journal = "Phys. Rev. D",
    volume = "104",
    number = "5",
    pages = "054016",
    year = "2021"
}

@article{Wang:2023iox,
    author = "Wang, Fu-Lai and Liu, Xiang",
    title = "{Higher molecular P{\ensuremath{\psi}}s{\ensuremath{\Lambda}}/{\ensuremath{\Sigma}} pentaquarks arising from the {\ensuremath{\Xi}}c(',*)D{\textasciimacron}1/{\ensuremath{\Xi}}c(',*)D{\textasciimacron}2* interactions}",
    eprint = "2307.08276",
    archivePrefix = "arXiv",
    primaryClass = "hep-ph",
    doi = "10.1103/PhysRevD.108.054028",
    journal = "Phys. Rev. D",
    volume = "108",
    number = "5",
    pages = "054028",
    year = "2023"
}

@article{Gao:2021hmv,
    author = "Gao, Feng and Li, Hao-Song",
    title = "{Magnetic moments of hidden-charm strange pentaquark states*}",
    eprint = "2112.01823",
    archivePrefix = "arXiv",
    primaryClass = "hep-ph",
    doi = "10.1088/1674-1137/ac8651",
    journal = "Chin. Phys. C",
    volume = "46",
    number = "12",
    pages = "123111",
    year = "2022"
}

@article{Guo:2023fih,
    author = "Guo, Fei and Li, Hao-Song",
    title = "{Analysis of the hidden-charm pentaquark states based on magnetic moment and transition magnetic moment}",
    eprint = "2304.10981",
    archivePrefix = "arXiv",
    primaryClass = "hep-ph",
    doi = "10.1140/epjc/s10052-024-12699-5",
    journal = "Eur. Phys. J. C",
    volume = "84",
    number = "4",
    pages = "392",
    year = "2024"
}

@article{Ozdem:2022iqk,
    author = {{\"O}zdem, Ula{\c{s}}},
    title = "{Magnetic moments of pentaquark states in light-cone sum rules}",
    doi = "10.1140/epja/s10050-022-00700-2",
    journal = "Eur. Phys. J. A",
    volume = "58",
    number = "3",
    pages = "46",
    year = "2022"
}

@article{Wang:2022nqs,
    author = "Wang, Fu-Lai and Luo, Si-Qiang and Zhou, Hong-Yan and Liu, Zhan-Wei and Liu, Xiang",
    title = "{Exploring the electromagnetic properties of the {\ensuremath{\Xi}}c(',*)D{\textasciimacron}s* and {\ensuremath{\Omega}}c(*)D{\textasciimacron}s* molecular states}",
    eprint = "2210.02809",
    archivePrefix = "arXiv",
    primaryClass = "hep-ph",
    doi = "10.1103/PhysRevD.108.034006",
    journal = "Phys. Rev. D",
    volume = "108",
    number = "3",
    pages = "034006",
    year = "2023"
}

@article{Wang:2022tib,
    author = "Wang, Fu-Lai and Zhou, Hong-Yan and Liu, Zhan-Wei and Liu, Xiang",
    title = "{What can we learn from the electromagnetic properties of hidden-charm molecular pentaquarks with single strangeness?}",
    eprint = "2208.10756",
    archivePrefix = "arXiv",
    primaryClass = "hep-ph",
    doi = "10.1103/PhysRevD.106.054020",
    journal = "Phys. Rev. D",
    volume = "106",
    number = "5",
    pages = "054020",
    year = "2022"
}

@article{Ozdem:2024jty,
    author = {{\"O}zdem, Ula{\c{s}}},
    title = "{Analysis of the isospin eigenstate $\bar{D} \Sigma _c$, $\bar{D}^{*} \Sigma _c$, and $\bar{D} \Sigma _c^{*}$ pentaquarks by their electromagnetic properties}",
    eprint = "2401.12678",
    archivePrefix = "arXiv",
    primaryClass = "hep-ph",
    doi = "10.1140/epjc/s10052-024-13124-7",
    journal = "Eur. Phys. J. C",
    volume = "84",
    number = "8",
    pages = "769",
    year = "2024"
}

@article{Li:2024wxr,
    author = "Li, Hao-Song and Guo, Fei and Lei, Ya-Ding and Gao, Feng",
    title = "{Magnetic moments and axial charges of the octet hidden-charm molecular pentaquark family}",
    eprint = "2401.14767",
    archivePrefix = "arXiv",
    primaryClass = "hep-ph",
    doi = "10.1103/PhysRevD.109.094027",
    journal = "Phys. Rev. D",
    volume = "109",
    number = "9",
    pages = "094027",
    year = "2024"
}

@article{Li:2024jlq,
    author = "Li, Hao-Song",
    title = "{Molecular pentaquark magnetic moments in heavy pentaquark chiral perturbation theory}",
    eprint = "2401.14759",
    archivePrefix = "arXiv",
    primaryClass = "hep-ph",
    doi = "10.1103/PhysRevD.109.114039",
    journal = "Phys. Rev. D",
    volume = "109",
    number = "11",
    pages = "114039",
    year = "2024"
}

@article{Ozdem:2024yel,
    author = {{\"O}zdem, Ula{\c{s}}},
    title = "{Investigation on the electromagnetic properties of the $ D^{(*)} \Sigma _c^{(*)}$ molecules}",
    eprint = "2405.07273",
    archivePrefix = "arXiv",
    primaryClass = "hep-ph",
    doi = "10.1140/epja/s10050-024-01477-2",
    journal = "Eur. Phys. J. A",
    volume = "61",
    number = "1",
    pages = "10",
    year = "2025"
}

@article{Mutuk:2024ltc,
    author = "Mutuk, Halil and Kang, Xian-Wei",
    title = "{Unveiling the structure of hidden-bottom strange pentaquarks via magnetic moments}",
    eprint = "2405.07066",
    archivePrefix = "arXiv",
    primaryClass = "hep-ph",
    doi = "10.1016/j.physletb.2024.138772",
    journal = "Phys. Lett. B",
    volume = "855",
    pages = "138772",
    year = "2024"
}

@article{Mutuk:2024jxf,
    author = "Mutuk, Halil",
    title = "{Magnetic moments of hidden-bottom pentaquark states}",
    eprint = "2403.16616",
    archivePrefix = "arXiv",
    primaryClass = "hep-ph",
    doi = "10.1140/epjc/s10052-024-13263-x",
    journal = "Eur. Phys. J. C",
    volume = "84",
    number = "8",
    pages = "874",
    year = "2024"
}

@article{Mutuk:2024ach,
    author = "Mutuk, Halil",
    title = "{Magnetic moments of hidden-charm pentaquarks in the diquark{\textendash}diquark{\textendash}antiquark scheme}",
    eprint = "2411.16486",
    archivePrefix = "arXiv",
    primaryClass = "hep-ph",
    doi = "10.1016/j.cjph.2025.07.030",
    journal = "Chin. J. Phys.",
    volume = "97",
    pages = "1406--1414",
    year = "2025"
}

@article{Ozdem:2025ncd,
    author = {{\"O}zdem, Ula{\c{s}}},
    title = "{Unveiling the electromagnetic structure and intrinsic dynamics of spin-3/2 hidden-charm pentaquarks: A comprehensive QCD analysis}",
    eprint = "2504.13488",
    archivePrefix = "arXiv",
    primaryClass = "hep-ph",
    doi = "10.1088/1674-1137/ade95a",
    journal = "Chin. Phys.",
    volume = "49",
    number = "10",
    pages = "103106",
    year = "2025"
}

@article{Ozdem:2025fks,
    author = {{\"O}zdem, Ula{\c{s}}},
    title = "{Probing the electromagnetic structure of the $P_c(4337)^+$ pentaquark: insights from a diquark{\textendash}diquark{\textendash}antiquark picture for $J^P = \frac{1}{2}^-$ and $\frac{3}{2}^-$ states}",
    eprint = "2506.04345",
    archivePrefix = "arXiv",
    primaryClass = "hep-ph",
    doi = "10.1140/epjc/s10052-025-14439-9",
    journal = "Eur. Phys. J. C",
    volume = "85",
    number = "6",
    pages = "704",
    year = "2025"
}

@article{Ozdem:2025jda,
    author = {{\"O}zdem, Ula{\c{s}}},
    title = "{Electromagnetic tomography of spin-$ \frac{3}{2} $ hidden-charm strange pentaquarks}",
    eprint = "2510.26893",
    archivePrefix = "arXiv",
    primaryClass = "hep-ph",
    doi = "10.1007/JHEP02(2026)207",
    journal = "JHEP",
    volume = "02",
    pages = "207",
    year = "2026"
}

@article{Chernyak:1990ag,
    author = "Chernyak, V. L. and Zhitnitsky, I. R.",
    title = "{B meson exclusive decays into baryons}",
    doi = "10.1016/0550-3213(90)90612-H",
    journal = "Nucl. Phys. B",
    volume = "345",
    pages = "137--172",
    year = "1990"
}

@article{Braun:1988qv,
    author = "Braun, Vladimir M. and Filyanov, I. E.",
    title = "{QCD Sum Rules in Exclusive Kinematics and Pion Wave Function}",
    reportNumber = "LENINGRAD-88-1446",
    doi = "10.1007/BF01548594",
    journal = "Z. Phys. C",
    volume = "44",
    pages = "157",
    year = "1989"
}

@article{Balitsky:1989ry,
    author = "Balitsky, I. I. and Braun, Vladimir M. and Kolesnichenko, A. V.",
    title = "{Radiative Decay Sigma+ ---{\ensuremath{>}} p gamma in Quantum Chromodynamics}",
    doi = "10.1016/0550-3213(89)90570-1",
    journal = "Nucl. Phys. B",
    volume = "312",
    pages = "509--550",
    year = "1989"
}

@article{Leinweber:1990dv,
    author = "Leinweber, Derek B. and Woloshyn, R. M. and Draper, Terrence",
    title = "{Electromagnetic structure of octet baryons}",
    reportNumber = "TRI-PP-90-52, UK-PP-90-09",
    doi = "10.1103/PhysRevD.43.1659",
    journal = "Phys. Rev. D",
    volume = "43",
    pages = "1659--1678",
    year = "1991"
}

@article{Weber:1978dh,
    author = "Weber, H. J. and Arenhovel, H.",
    title = "{Isobar Configurations in Nuclei}",
    doi = "10.1016/0370-1573(78)90187-4",
    journal = "Phys. Rept.",
    volume = "36",
    pages = "277--348",
    year = "1978"
}

@article{Nozawa:1990gt,
    author = "Nozawa, S. and Leinweber, D. B.",
    title = "{Electromagnetic form-factors of spin 3/2 baryons}",
    reportNumber = "TRI-PP-90-19",
    doi = "10.1103/PhysRevD.42.3567",
    journal = "Phys. Rev. D",
    volume = "42",
    pages = "3567--3571",
    year = "1990"
}

@article{Pascalutsa:2006up,
    author = "Pascalutsa, Vladimir and Vanderhaeghen, Marc and Yang, Shin Nan",
    title = "{Electromagnetic excitation of the Delta(1232)-resonance}",
    eprint = "hep-ph/0609004",
    archivePrefix = "arXiv",
    reportNumber = "JLAB-THY-06-537",
    doi = "10.1016/j.physrep.2006.09.006",
    journal = "Phys. Rept.",
    volume = "437",
    pages = "125--232",
    year = "2007"
}

@article{Ramalho:2009vc,
    author = "Ramalho, G. and Pena, M. T. and Gross, Franz",
    title = "{Electric quadrupole and magnetic octupole moments of the Delta}",
    eprint = "0902.4212",
    archivePrefix = "arXiv",
    primaryClass = "hep-ph",
    reportNumber = "JLAB-THY-09-951",
    doi = "10.1016/j.physletb.2009.06.052",
    journal = "Phys. Lett. B",
    volume = "678",
    pages = "355--358",
    year = "2009"
}

@article{Balitsky:1987bk,
    author = "Balitsky, I. I. and Braun, Vladimir M.",
    title = "{Evolution Equations for QCD String Operators}",
    reportNumber = "LENINGRAD-87-1351",
    doi = "10.1016/0550-3213(89)90168-5",
    journal = "Nucl. Phys. B",
    volume = "311",
    pages = "541--584",
    year = "1989"
}

@article{Belyaev:1985wza,
    author = "Belyaev, V. M. and Blok, B. Yu.",
    title = "{CHARMED BARYONS IN QUANTUM CHROMODYNAMICS}",
    doi = "10.1007/BF01560689",
    journal = "Z. Phys. C",
    volume = "30",
    pages = "151",
    year = "1986"
}

@article{Ball:2002ps,
    author = "Ball, Patricia and Braun, V. M. and Kivel, N.",
    title = "{Photon distribution amplitudes in QCD}",
    eprint = "hep-ph/0207307",
    archivePrefix = "arXiv",
    reportNumber = "IPPP-02-40, DCPT-02-80",
    doi = "10.1016/S0550-3213(02)01017-9",
    journal = "Nucl. Phys. B",
    volume = "649",
    pages = "263--296",
    year = "2003"
}

@article{Workman:2022ynf,
    author = "Workman, R. L. and others",
    collaboration = "Particle Data Group",
    title = "{Review of Particle Physics}",
    doi = "10.1093/ptep/ptac097",
    journal = "PTEP",
    volume = "2022",
    pages = "083C01",
    year = "2022"
}

@article{Ioffe:2005ym,
    author = "Ioffe, B. L.",
    title = "{QCD at low energies}",
    eprint = "hep-ph/0502148",
    archivePrefix = "arXiv",
    doi = "10.1016/j.ppnp.2005.05.001",
    journal = "Prog. Part. Nucl. Phys.",
    volume = "56",
    pages = "232--277",
    year = "2006"
}

@article{Narison:2018nbv,
    author = "Narison, Stephan",
    editor = "Narison, St{\'e}phan",
    title = "{$\overline{\rm m}_{c,b,}<\alpha_sG^2>$ and $\alpha_s$ from Heavy Quarkonia}",
    doi = "10.1016/j.nuclphysbps.2018.12.026",
    journal = "Nucl. Part. Phys. Proc.",
    volume = "300-302",
    pages = "153--164",
    year = "2018"
}

@article{Rohrwild:2007yt,
    author = "Rohrwild, J.",
    title = "{Determination of the magnetic susceptibility of the quark condensate using radiative heavy meson decays}",
    eprint = "0708.1405",
    archivePrefix = "arXiv",
    primaryClass = "hep-ph",
    doi = "10.1088/1126-6708/2007/09/073",
    journal = "JHEP",
    volume = "09",
    pages = "073",
    year = "2007"
}

@article{Zakharov:1968fb,
    author = "Zakharov, V. I. and Kondratyuk, L. A. and Ponomarev, L. A.",
    title = "{Bremsstrahlung and determination of electromagnetic parameters of particles}",
    journal = "Yad. Fiz.",
    volume = "8",
    pages = "783--792",
    year = "1968"
}

@article{Pascalutsa:2004je,
    author = "Pascalutsa, Vladimir and Vanderhaeghen, Marc",
    title = "{Magnetic moment of the Delta(1232)-resonance in chiral effective field theory}",
    eprint = "nucl-th/0412113",
    archivePrefix = "arXiv",
    reportNumber = "WM-04-124, JLAB-THY-05-292",
    doi = "10.1103/PhysRevLett.94.102003",
    journal = "Phys. Rev. Lett.",
    volume = "94",
    pages = "102003",
    year = "2005"
}

@article{Ozdem:2024txt,
    author = {{\"O}zdem, Ula{\c{s}}},
    title = "{Investigating the underlying structure of vector hidden-charm tetraquark states via their electromagnetic characteristics}",
    eprint = "2412.06447",
    archivePrefix = "arXiv",
    primaryClass = "hep-ph",
    doi = "10.1103/PhysRevD.111.054009",
    journal = "Phys. Rev. D",
    volume = "111",
    number = "5",
    pages = "054009",
    year = "2025"
}

@article{Ozdem:2024dbq,
    author = {{\"O}zdem, Ula{\c{s}}},
    title = "{Unveiling the underlying structure of axial-vector bottom-charm tetraquarks in the light of their magnetic moments}",
    eprint = "2403.16191",
    archivePrefix = "arXiv",
    primaryClass = "hep-ph",
    doi = "10.1007/JHEP05(2024)301",
    journal = "JHEP",
    volume = "05",
    pages = "301",
    year = "2024"
}

@article{Azizi:2023gzv,
    author = {Azizi, K. and {\"O}zdem, U.},
    title = "{Exploring the magnetic dipole moments of $ {T}_{QQ\overline{q}\overline{s}} $ and $ {T}_{QQ\overline{s}\overline{s}} $ states in the framework of QCD light-cone sum rules}",
    eprint = "2301.07713",
    archivePrefix = "arXiv",
    primaryClass = "hep-ph",
    doi = "10.1007/JHEP03(2023)166",
    journal = "JHEP",
    volume = "03",
    pages = "166",
    year = "2023"
}

@article{Ozdem:2024lpk,
    author = {{\"O}zdem, U. and Azizi, K.},
    title = "{Electromagnetic properties of vector doubly charmed tetraquark states}",
    eprint = "2401.04798",
    archivePrefix = "arXiv",
    primaryClass = "hep-ph",
    doi = "10.1103/PhysRevD.109.114019",
    journal = "Phys. Rev. D",
    volume = "109",
    number = "11",
    pages = "114019",
    year = "2024"
}

@article{Ozdem:2024suc,
    author = {{\"O}zdem, Ula{\c{s}}},
    title = "{Insight into the nature of the $P_{c}(4457)$ and related pentaquarks}",
    eprint = "2409.09449",
    archivePrefix = "arXiv",
    primaryClass = "hep-ph",
    doi = "10.1140/epjc/s10052-025-14323-6",
    journal = "Eur. Phys. J. C",
    volume = "85",
    number = "6",
    pages = "624",
    year = "2025"
}

\end{document}